\input harvmac.tex


\def\unlockat{\catcode`\@=11}

\def\lockat{\catcode`\@=12}

\unlockat


\def\newsec#1{\global\advance\secno by1\message{(\the\secno. #1)}
\global\subsecno=0\global\subsubsecno=0
\global\deno=0\global\prono=0\global\teno=0\eqnres@t\noindent
{\bf\the\secno. #1}
\writetoca{{\secsym} {#1}}\par\nobreak\medskip\nobreak}
\global\newcount\subsecno \global\subsecno=0
\def\subsec#1{\global\advance\subsecno by1\message{(\secsym\the\subsecno. #1)}
\ifnum\lastpenalty>9000\else\bigbreak\fi\global\subsubsecno=0
\global\deno=0\global\prono=0\global\teno=0
\noindent{\it\secsym\the\subsecno. #1}
\writetoca{\string\quad {\secsym\the\subsecno.} {#1}}
\par\nobreak\medskip\nobreak}
\global\newcount\subsubsecno \global\subsubsecno=0
\def\subsubsec#1{\global\advance\subsubsecno by1
\message{(\secsym\the\subsecno.\the\subsubsecno. #1)}
\ifnum\lastpenalty>9000\else\bigbreak\fi
\noindent\quad{\secsym\the\subsecno.\the\subsubsecno.}{#1}
\writetoca{\string\qquad{\secsym\the\subsecno.\the\subsubsecno.}{#1}}
\par\nobreak\medskip\nobreak}

\global\newcount\deno \global\deno=0
\def\de#1{\global\advance\deno by1
\message{(\bf Definition\quad\secsym\the\subsecno.\the\deno #1)}
\ifnum\lastpenalty>9000\else\bigbreak\fi
\noindent{\bf Definition\quad\secsym\the\subsecno.\the\deno}{#1}
\writetoca{\string\qquad{\secsym\the\subsecno.\the\deno}{#1}}}

\global\newcount\prono \global\prono=0
\def\pro#1{\global\advance\prono by1
\message{(\bf Proposition\quad\secsym\the\subsecno.\the\prono #1)}
\ifnum\lastpenalty>9000\else\bigbreak\fi
\noindent{\bf Proposition\quad\secsym\the\subsecno.\the\prono}{#1}
\writetoca{\string\qquad{\secsym\the\subsecno.\the\prono}{#1}}}

\global\newcount\teno \global\prono=0
\def\te#1{\global\advance\teno by1
\message{(\bf Theorem\quad\secsym\the\subsecno.\the\teno #1)}
\ifnum\lastpenalty>9000\else\bigbreak\fi
\noindent{\bf Theorem\quad\secsym\the\subsecno.\the\teno}{#1}
\writetoca{\string\qquad{\secsym\the\subsecno.\the\teno}{#1}}}
\def\subsubseclab#1{\DefWarn#1\xdef #1{\noexpand\hyperref{}{subsubsection}%
{\secsym\the\subsecno.\the\subsubsecno}%
{\secsym\the\subsecno.\the\subsubsecno}}%
\writedef{#1\leftbracket#1}\wrlabeL{#1=#1}}

\lockat

\def\IB{\relax\hbox{$\inbar\kern-.3em{\rm B}$}}
\def\IC{\relax\hbox{$\inbar\kern-.3em{\rm C}$}}
\def\ID{\relax\hbox{$\inbar\kern-.3em{\rm D}$}}
\def\IE{\relax\hbox{$\inbar\kern-.3em{\rm E}$}}
\def\IF{\relax\hbox{$\inbar\kern-.3em{\rm F}$}}
\def\IG{\relax\hbox{$\inbar\kern-.3em{\rm G}$}}
\def\IGa{\relax\hbox{${\rm I}\kern-.18em\Gamma$}}
\def\IH{\relax{\rm I\kern-.18em H}}
\def\IK{\relax{\rm I\kern-.18em K}}
\def\IL{\relax{\rm I\kern-.18em L}}
\def\IP{\relax{\rm I\kern-.18em P}}
\def\IR{\relax{\rm I\kern-.18em R}}
\def\IZ{\relax\ifmmode\mathchoice
{\hbox{\cmss Z\kern-.4em Z}}{\hbox{\cmss Z\kern-.4em Z}}
{\lower.9pt\hbox{\cmsss Z\kern-.4em Z}}
{\lower1.2pt\hbox{\cmsss Z\kern-.4em Z}}\else{\cmss Z\kern-.4em Z}\fi}

\def\II{\relax{\rm I\kern-.18em I}}

\def\CA {{\cal A}}

\def\CD {{\cal D}}

\def\CF {{\cal F}}
\def\CG {{\cal G}}

\def\CL {{\cal L}}
\def\CM {{\cal M}}
\def\CN {{\cal N}}
\def\CO {{\cal O}}

\def\CQ {{\cal Q}}
\def\CR {{\cal R}}
\def\CS {{\cal S}}
\def\CT {{\cal T}}

\def\CV {{\cal V}}

\def\CX {{\cal X}}


\def\p{\partial}
\def\pb{\bar{\partial}}

\def\wb {\bar{w}}
\def\zb {\bar{z}}


\def\Tr{{\rm Tr}}

\def\vol{{\rm vol}}
\def\Vol{{\rm Vol}}
\def\wzwt{$WZW_2$}
\def\wzwf{$WZW_4$}

\def\c{\cdot}
\def\sdtimes{\mathbin{\hbox{\hskip2pt\vrule height 4.1pt depth -.3pt
width
.25pt
\hskip-2pt$\times$}}}
\def\p{\partial}
\def\pb{\bar{\partial}}

\def\ch{{\rm ch}}
\def\Det{{\rm Det}}

\def\lieg{{\underline{\bf g}}}
\def\liet{{\underline{\bf t}}}
\def\liek{{\underline{\bf k}}}

\def\lieh{{\underline{\bf h}}}

\def\cliet{{\underline{\bf t}}_{\scriptstyle{\IC}}}
\def\cliek{{\underline{\bf k}}_{\scriptscriptstyle{\IC}}}

\def\CCK{K_{\scriptscriptstyle{\IC}}}
\def\inbar{\,\vrule height1.5ex width.4pt depth0pt}
\font\cmss=cmss10 \font\cmsss=cmss10 at 7pt
\def\sdtimes{\mathbin{\hbox{\hskip2pt\vrule
height 4.1pt depth -.3pt width .25pt\hskip-2pt$\times$}}}


\font\manual=manfnt \def\dbend{\lower3.5pt\hbox{\manual\char127}}


\def\boxit#1{\vbox{\hrule\hbox{\vrule\kern8pt
\vbox{\hbox{\kern8pt}\hbox{\vbox{#1}}\hbox{\kern8pt}}
\kern8pt\vrule}\hrule}}
\def\mathboxit#1{\vbox{\hrule\hbox{\vrule\kern8pt\vbox{\kern8pt
\hbox{$\displaystyle #1$}\kern8pt}\kern8pt\vrule}\hrule}}

%

\lref\gerasimovshatashvili{A.~Gerasimov, S.~Shatashvili, ``Higgs Bundles, Gauge Theories and Quantum Groups'', hep-th/0609024 }
\lref\nikthesis{N.~Nekrasov, ``Four dimensional holomorphic theories'',
PhD. thesis, Princeton University, 1996, UMI-9701221}


\lref\blzh{A. Belavin, V. Zakharov, ``Yang-Mills Equations as inverse
scattering
problem''Phys. Lett. B73, (1978) 53}
\lref\bost{L. Alvarez-Gaume, J.B. Bost , G. Moore, P. Nelson, C.
Vafa,
``Bosonization on higher genus Riemann surfaces,''
Commun.Math.Phys.112:503,1987}
\lref\agmv{L. Alvarez-Gaum\'e,
C. Gomez, G. Moore,
and C. Vafa, ``Strings in the Operator Formalism,''
Nucl. Phys. {\bf 303}(1988)455}
\lref\atiyah{M. Atiyah, ``Green's Functions for
Self-Dual Four-Manifolds,'' Adv. Math. Suppl.
{\bf 7A} (1981)129}

\lref\donagi{R.~Y.~ Donagi,
``Seiberg-Witten integrable systems'',
alg-geom/9705010 }

\lref\AHS{M.~ Atiyah, N.~ Hitchin and I.~ Singer, ``Self-Duality in
Four-Dimensional
Riemannian Geometry", Proc. Royal Soc. (London) {\bf A362} (1978)
425-461.}
\lref\fmlies{M. F. Atiyah and I. M. Singer,
``The index of elliptic operators IV,'' Ann. Math. {\bf 93}(1968)119}
\lref\BlThlgt{M.~ Blau and G.~ Thompson, ``Lectures on 2d Gauge
Theories: Topological Aspects and Path
Integral Techniques", Presented at the
Summer School in Hogh Energy Physics and
Cosmology, Trieste, Italy, 14 Jun - 30 Jul
1993, hep-th/9310144.}
\lref\bpz{A.A. Belavin, A.M. Polyakov, A.B. Zamolodchikov,
``Infinite conformal symmetry in two-dimensional quantum
field theory,'' Nucl.Phys.B241:333,1984}
\lref\braam{P.J. Braam, A. Maciocia, and A. Todorov,
``Instanton moduli as a novel map from tori to
K3-surfaces,'' Inven. Math. {\bf 108} (1992) 419}
\lref\CMR{ For a review, see
S. Cordes, G. Moore, and S. Ramgoolam,
`` Lectures on 2D Yang Mills theory, Equivariant
Cohomology, and Topological String Theory,''
Lectures presented at the 1994 Les Houches Summer School
 ``Fluctuating Geometries in Statistical Mechanics and Field
Theory.''
and at the Trieste 1994 Spring school on superstrings.
hep-th/9411210, or see http://xxx.lanl.gov/lh94}
\lref\dnld{S. Donaldson, ``Anti self-dual Yang-Mills
connections over complex  algebraic surfaces and stable
vector bundles,'' Proc. Lond. Math. Soc,
{\bf 50} (1985)1}

\lref\DoKro{S.K.~ Donaldson and P.B.~ Kronheimer,
{\it The Geometry of Four-Manifolds},
Clarendon Press, Oxford, 1990.}
\lref\donii{
S. Donaldson, Duke Math. J. , {\bf 54} (1987) 231. }

\lref\fs{L. Faddeev and S. Shatashvili, Theor. Math. Fiz., 60 (1984)
206}
\lref\fsi{ L. Faddeev, Phys. Lett. B145 (1984) 81.}
\lref\fadba{L.D.~Faddeev, ``How  algebraic Bethe ansatz works for
integrable
model'',  hep-th/9605187}
\lref\fadbai{L.D.~Faddeev,
``Algebraic aspects of Bethe ansatz'',  Int.J.Mod.Phys.{\bf A}10 (1995)
1845-1878,
 hep-th/9404013 \semi
``The Bethe  ansatz'', SFB-288-70, Jun 1993. Andrejewski lectures}
\lref\faddeevlmp{L. D. Faddeev, ``Some Comments on Many Dimensional Solitons'',
Lett. Math. Phys., 1 (1976) 289-293.}

\lref\gerasimov{A. Gerasimov, ``Localization in
GWZW and Verlinde formula,'' hepth/9305090}

\lref\gottsh{L. Gottsche, Math. Ann. 286 (1990)193}
\lref\GrHa{P.~ Griffiths and J.~ Harris, {\it Principles of
Algebraic
geometry},
p. 445, J.Wiley and Sons, 1978. }

\lref\hitchin{N. Hitchin, ``Polygons and gravitons,''
Math. Proc. Camb. Phil. Soc, (1979){\bf 85} 465}
\lref\hi{N.~Hitchin, ``Stable bundles and integrable systems'', Duke Math
{\bf 54}  (1987),91-114}
\lref\hid{N.~Hitchin, ``The self-duality equations on a Riemann surface'',
Proc. London Math. Soc. {\bf 55} (1987) 59-126 }
\lref\hklr{N.~Hitchin, A.~Karlhede, U.~Lindstrom, and M.~Rocek,
``Hyperkahler metrics and supersymmetry,''
Commun. Math. Phys. {\bf 108}(1987)535}
\lref\hirz{F. Hirzebruch and T. Hofer, Math. Ann. 286 (1990)255}
\lref\btverlinde{M.~ Blau, G.~ Thomson,
``Derivation of the Verlinde Formula from Chern-Simons Theory and the
$G/G$
   model'',Nucl. Phys. {\bf B}408 (1993) 345-390 }
\lref\kronheimer{P. Kronheimer, ``The construction of ALE spaces as
hyper-kahler quotients,'' J. Diff. Geom. {\bf 28}1989)665}
\lref\kricm{P. Kronheimer, ``Embedded surfaces in
4-manifolds,'' Proc. Int. Cong. of
Math. (Kyoto 1990) ed. I. Satake, Tokyo, 1991}

\lref\krmw{P.~Kronheimer and T.~Mrowka,
``Gauge theories for embedded surfaces I,''
Topology {\bf 32} (1993) 773,
``Gauge theories for embedded surfaces II,''
preprint.}
\lref\kirwan{F.~Kirwan, ``Cohomology of quotients in symplectic
and algebraic geometry'', Math. Notes, Princeton University Press, 1985}
\lref\avatar{A. Losev, G. Moore, N. Nekrasov, S. Shatashvili,
``Four-Dimensional Avatars of 2D RCFT,''
hep-th/9509151, Nucl.Phys.Proc.Suppl.46:130-145,1996 }
\lref\cocycle{A. Losev, G. Moore, N. Nekrasov, S. Shatashvili,
``Central Extensions of Gauge Groups Revisited,''
hep-th/9511185.}
\lref\maciocia{A. Maciocia, ``Metrics on the moduli
spaces of instantons over Euclidean 4-Space,''
Commun. Math. Phys. {\bf 135}(1991) , 467}
\lref\mickold{J. Mickelsson, CMP, 97 (1985) 361.}
\lref\mick{J. Mickelsson, ``Kac-Moody groups,
topology of the Dirac determinant bundle and
fermionization,'' Commun. Math. Phys., {\bf 110} (1987) 173.}
\lref\milnor{J. Milnor, ``A unique decomposition
theorem for 3-manifolds,'' Amer. Jour. Math, (1961) 1}
\lref\taming{G. Moore and N. Seiberg,
``Taming the conformal zoo,'' Phys. Lett.
{\bf 220 B} (1989) 422}
\lref\nair{V.P.Nair, ``K\"ahler-Chern-Simons Theory'', hep-th/9110042}
\lref\ns{V.P. Nair and Jeremy Schiff,
``Kahler Chern Simons theory and symmetries of
antiselfdual equations'' Nucl.Phys.B371:329-352,1992;
``A Kahler Chern-Simons theory and quantization of the
moduli of antiselfdual instantons,''
Phys.Lett.B246:423-429,1990,
``Topological gauge theory and twistors,''
Phys.Lett.B233:343,1989}
\lref\ogvf{H. Ooguri and C. Vafa, ``Self-Duality
and $N=2$ String Magic,'' Mod.Phys.Lett. {\bf A5} (1990) 1389-1398;
``Geometry
of$N=2$ Strings,'' Nucl.Phys. {\bf B361}  (1991) 469-518.}
\lref\park{J.-S. Park, ``Holomorphic Yang-Mills theory on compact
Kahler
manifolds,'' hep-th/9305095; Nucl. Phys. {\bf B423} (1994) 559;
J.-S.~ Park, ``$N=2$ Topological Yang-Mills Theory on Compact
K\"ahler
Surfaces", Commun. Math, Phys. {\bf 163} (1994) 113;
S. Hyun and J.-S.~ Park, ``$N=2$ Topological Yang-Mills Theories and Donaldson
Polynomials", hep-th/9404009}
\lref\parki{S. Hyun and J.-S. Park,
``Holomorphic Yang-Mills Theory and Variation
of the Donaldson Invariants,'' hep-th/9503036}
\lref\dpark{J.-S.~Park, ``Monads and D-instantons'', hep-th/9612096}
\lref\pohl{Pohlmeyer, Commun.
Math. Phys. {\bf 72}(1980)37}
\lref\pwf{A.M. Polyakov and P.B. Wiegmann,
Phys. Lett. {\bf B131}(1983)121}
\lref\clash{
A.~Losev, G.~Moore, N.~Nekrasov, S.~Shatashvili,
`` Chiral  Lagrangians, Anomalies, Supersymmetry, and Holomorphy'', Nucl.Phys.
{\bf B} 484(1997) 196-222, hep-th/9606082 }
\lref\givental{A.B.~Givental,
``Equivariant Gromov - Witten Invariants'',
alg-geom/9603021}
\lref\prseg{Pressley and Segal, Loop Groups}
\lref\rade{J. Rade, ``Singular Yang-Mills fields. Local
theory I. '' J. reine ang. Math. , {\bf 452}(1994)111; {\it ibid}
{\bf 456}(1994)197; ``Singular Yang-Mills
fields-global theory,'' Intl. J. of Math. {\bf 5}(1994)491.}
\lref\segal{G. Segal, The definition of CFT}
\lref\sen{A. Sen,
hep-th/9402032, Dyon-Monopole bound states, selfdual harmonic
forms on the multimonopole moduli space and $SL(2,Z)$
invariance,'' }
\lref\shatashi{S. Shatashvili,
Theor. and Math. Physics, 71, 1987, p. 366}
\lref\thooft{G. 't Hooft , ``A property of electric and
magnetic flux in nonabelian gauge theories,''
Nucl.Phys.B153:141,1979}
\lref\vafa{C. Vafa, ``Conformal theories and punctured
surfaces,'' Phys.Lett.199B:195,1987 }
\lref\vrlsq{E. Verlinde and H. Verlinde,
``Conformal Field Theory and Geometric Quantization,''
in {\it Strings'89},Proceedings
of the Trieste Spring School on Superstrings,
3-14 April 1989, M. Green, et. al. Eds. World
Scientific, 1990}

\lref\mwxllvrld{E. Verlinde, ``Global Aspects of
Electric-Magnetic Duality,'' hep-th/9506011}

\lref\wrdhd{R. Ward, Nucl. Phys. {\bf B236}(1984)381}
\lref\ward{Ward and Wells, {\it Twistor Geometry and
Field Theory}, CUP }


\lref\WitDonagi{R.~ Donagi, E.~ Witten,
``Supersymmetric Yang-Mills Theory and
Integrable Systems'', hep-th/9510101, Nucl.Phys.{\bf B}460 (1996) 299-334}
\lref\Witfeb{E.~ Witten, ``Supersymmetric Yang-Mills Theory On A
Four-Manifold,'' J. Math. Phys. {\bf 35} (1994) 5101.}
\lref\Witr{E.~ Witten, ``Introduction to Cohomological Field
Theories",
Lectures at Workshop on Topological Methods in Physics, Trieste, Italy,
Jun 11-25, 1990, Int. J. Mod. Phys. {\bf A6} (1991) 2775.}
\lref\Witgrav{E.~ Witten, ``Topological Gravity'', Phys.Lett.206B:601, 1988}
\lref\witaffl{I. ~ Affleck, J.A.~ Harvey and E.~ Witten,
    ``Instantons and (Super)Symmetry Breaking
    in $2+1$ Dimensions'', Nucl. Phys. {\bf B}206 (1982) 413}
\lref\wittabl{E.~ Witten,  ``On $S$-Duality in Abelian Gauge Theory,''
hep-th/9505186; Selecta Mathematica {\bf 1} (1995) 383}
\lref\wittgr{E.~ Witten, ``The Verlinde Algebra And The Cohomology Of
The Grassmannian'',  hep-th/9312104}
\lref\wittenwzw{E. Witten, ``Nonabelian bosonization in
two dimensions,'' Commun. Math. Phys. {\bf 92} (1984)455 }
\lref\witgrsm{E. Witten, ``Quantum field theory,
grassmannians and algebraic curves,'' Commun.Math.Phys.113:529,1988}
\lref\wittjones{E. Witten, ``Quantum field theory and the Jones
polynomial,'' Commun.  Math. Phys., 121 (1989) 351. }
\lref\witttft{E.~ Witten, ``Topological Quantum Field Theory",
Commun. Math. Phys. {\bf 117} (1988) 353.}
\lref\wittmon{E.~ Witten, ``Monopoles and Four-Manifolds'', hep-th/9411102}
\lref\Witdgt{ E.~ Witten, ``On Quantum gauge theories in two
dimensions,''
Commun. Math. Phys. {\bf  141}  (1991) 153\semi
 ``Two dimensional gauge
theories revisited'', J. Geom. Phys. 9 (1992) 303-368}
\lref\Witgenus{E.~ Witten, ``Elliptic Genera and Quantum Field Theory'',
Comm. Math. Phys. 109(1987) 525. }
\lref\OldZT{E. Witten, ``New Issues in Manifolds of SU(3) Holonomy,''
{\it Nucl. Phys.} {\bf B268} (1986) 79 \semi
J. Distler and B. Greene, ``Aspects of (2,0) String Compactifications,''
{\it Nucl. Phys.} {\bf B304} (1988) 1 \semi
B. Greene, ``Superconformal Compactifications in Weighted Projective
Space,'' {\it Comm. Math. Phys.} {\bf 130} (1990) 335.}

\lref\bagger{E.~ Witten and J. Bagger, Phys. Lett.
{\bf 115B}(1982) 202}

\lref\witcurrent{E.~ Witten,``Global Aspects of Current Algebra'',
Nucl.Phys.B223 (1983) 422\semi
``Current Algebra, Baryons and Quark Confinement'', Nucl.Phys. B223 (1993)
433}
\lref\Wittreiman{S.B. Treiman,
E. Witten, R. Jackiw, B. Zumino, ``Current Algebra and
Anomalies'', Singapore, Singapore: World Scientific ( 1985) }
\lref\Witgravanom{L. Alvarez-Gaume, E.~ Witten, ``Gravitational Anomalies'',
Nucl.Phys.B234:269,1984. }

\lref\CHSW{P.~Candelas, G.~Horowitz, A.~Strominger and E.~Witten,
``Vacuum Configurations for Superstrings,'' {\it Nucl. Phys.} {\bf
B258} (1985) 46.}

\lref\AandB{E.~Witten, in ``Proceedings of the Conference on Mirror Symmetry",
MSRI (1991).}

\lref\phases{E.~Witten, ``Phases of N=2 Theories in Two Dimensions",
Nucl. Phys. {\bf B403} (1993) 159, hep-th/9301042}
\lref\WitKachru{S.~Kachru and E.~Witten, ``Computing The Complete Massless
Spectrum Of A Landau-Ginzburg Orbifold,''
Nucl. Phys. {\bf B407} (1993) 637, hep-th/9307038}

\lref\WitMin{E.~Witten,
``On the Landau-Ginzburg Description of N=2 Minimal Models,''
IASSNS-HEP-93/10, hep-th/9304026.}

\lref\wittenwzw{E. Witten, ``Nonabelian bosonization in
two dimensions,'' Commun. Math. Phys. {\bf 92} (1984)455 }
\lref\grssmm{E. Witten, ``Quantum field theory,
grassmannians and algebraic curves,'' Commun.Math.Phys.113:529,1988}
\lref\wittjones{E. Witten, ``Quantum field theory and the Jones
polynomial,'' Commun.  Math. Phys., 121 (1989) 351. }
\lref\wittentft{E.~ Witten, ``Topological Quantum Field Theory",
Commun. Math. Phys. {\bf 117} (1988) 353.}
\lref\Witdgt{ E.~ Witten, ``On Quantum gauge theories in two
dimensions,''
Commun. Math. Phys. {\bf  141}  (1991) 153.}
\lref\Witfeb{E.~ Witten, ``Supersymmetric Yang-Mills Theory On A
Four-Manifold,'' J. Math. Phys. {\bf 35} (1994) 5101.}
\lref\Witr{E.~ Witten, ``Introduction to Cohomological Field
Theories",
Lectures at Workshop on Topological Methods in Physics, Trieste,
Italy,
Jun 11-25, 1990, Int. J. Mod. Phys. {\bf A6} (1991) 2775.}
\lref\wittabl{E. Witten,  ``On S-Duality in Abelian Gauge Theory,''
hep-th/9505186}

\lref\seiken{K. Intriligator, N. Seiberg,
``Mirror Symmetry in Three Dimensional Gauge Theories'',
hep-th/9607207, Phys.Lett. B387 (1996) 513}
\lref\douglas{M.R. Douglas, ``Enhanced Gauge
Symmetry in M(atrix) Theory,'' hep-th/9612126}
\lref\hs{J.A. Harvey and A. Strominger,
``The heterotic string is a soliton,''
hep-th/9504047}
\lref\hm{ J.A.~Harvey, G.~Moore,
``On the algebras of BPS states'', hep-th/9609017}
\lref\zt{O.~Aharony, M.~Berkooz, N.~Seiberg, ``Light-cone description
of $(2,0)$ supersonformal theories in six dimensions'', hep-th/9712117}
\lref\sen{A. Sen, `` String- String Duality Conjecture In Six Dimensions And
Charged Solitonic Strings'',  hep-th/9504027}
\lref\KN{P.~Kronheimer and H.~Nakajima,  ``Yang-Mills instantons
on ALE gravitational instantons,''  Math. Ann.
{\bf 288}(1990)263}
\lref\nakajima{H.~Nakajima, ``Homology of moduli
spaces of instantons on ALE Spaces. I'' J. Diff. Geom.
{\bf 40}(1990) 105; ``Instantons on ALE spaces,
quiver varieties, and Kac-Moody algebras,'' Duke. Math. J. {\bf 76} (1994)
365-416\semi
``Gauge theory on resolutions of simple singularities
and affine Lie algebras,'' Inter. Math. Res. Notices (1994), 61-74}
\lref\nakheis{H.~Nakajima, ``Heisenberg algebra and Hilbert schemes of
points on
projective surfaces ,'' alg-geom/9507012\semi
``Lectures on Hilbert schemes of points on surfaces'', H.~Nakajima's
homepage}
\lref\vw{C.~Vafa, E.~Witten, ``A strong coupling test of $S$-duality'',
Nucl. Phys. {\bf B} 431 (1994) 3-77}
\lref\grojn{I. Grojnowski, ``Instantons and
affine algebras I: the Hilbert scheme and
vertex operators,'' alg-geom/9506020}
\lref\gr{G.~Gibbons, P.~Rychenkova ``hyperkahler quotient construction
of BPS Monopole moduli space'', hep-th/9608085}
\lref\dvafa{
C.~Vafa, ``Instantons on D-branes'', hep-th/9512078,
Nucl.Phys. B463 (1996) 435-442}
\lref\atbott{M.~Atiyah, R.~Bott, ``The Moment Map And
Equivariant Cohomology'', Topology {\bf 23} (1984) 1-28}
\lref\atbotti{M.~Atiyah, R.~Bott, ``The Yang-Mills Equations Over
Riemann Surfaces'', Phil. Trans. R.Soc. London A {\bf 308}, 523-615 (1982)}

\Title{ \vbox{\baselineskip12pt\hbox{hep-th/9712241}
\hbox{ITEP-TH.62/97}
\hbox{HUTP- 97/A089}
\hbox{YCTP-P23/97}}}
{\vbox{
 \centerline{INTEGRATING OVER HIGGS BRANCHES}}}
\medskip
\centerline{Gregory Moore $^1$,
Nikita Nekrasov $^2$, and Samson Shatashvili $^{3}$\footnote{*}{On
leave of
absence from St. Petersburg Branch of Steklov Mathematical Institute, Fontanka,
St.
Petersburg,
Russia.}}

\vskip 0.5cm
\centerline{$^{2}$ Institute of Theoretical and Experimental
Physics,
117259, Moscow, Russia}
\centerline{$^2$ Lyman Laboratory of Physics,
Harvard University, Cambridge, MA 02138}
\centerline{$^{1,3}$ Department of Physics, Yale University,
New Haven, CT  06520, Box 208120}
\vskip 0.1cm
\centerline{moore@castalia.physics.yale.edu}
\centerline{nikita@string.harvard.edu }
\centerline{samson@euler.physics.yale.edu}
\centerline{\footnote{**}{in the year 2006: gmoore@physics.rutgers.edu, nikita@ihes.fr, samson@maths.tcd.ie}}
\medskip
\noindent
We develop some useful techinques for integrating over
Higgs branches in supersymmetric theories
with $4$  and $8$ supercharges. In
particular,
we define a regularized volume for hyperkahler
quotients. We evaluate this volume for certain
ALE and ALF spaces
in terms of the hyperkahler periods. We also reduce these
volumes for a large class of hyperkahler quotients to  simpler integrals.
These quotients
include complex coadjoint orbits, instanton  moduli spaces on
$\IR^{4}$ and ALE manifolds, Hitchin
spaces, and
moduli spaces of (parabolic) Higgs bundles on Riemann surfaces.
In the case of Hitchin spaces
the evaluation of the
volume reduces to a summation over solutions
of Bethe Ansatz equations for the
non-linear Schr\"odinger system.
We discuss some applications
of our results.

\Date{Dec. 27, 1997}

\newsec{Introduction }

In this note we study integrals over Higgs branches. They are
kahler and hyperkahler quotients, depending on the amount of
supersymmetry in the theory.

Our original motivation was inspired by the works \vw\ on $S$-duality
and \nakajima\ on the action of affine algebras on the moduli spaces
of instantons
which have been subsequently developed in many ways \grojn\avatar\nakheis.
In particular, the series of papers \avatar\cocycle\clash\ arose
out of attempts to give a field-theoretic interpretation of
the results of \nakajima. Subsequently,   string duality ideas turned
out to be more adequate for addressing these questions \dvafa\hm\ although
in all cases (\avatar\dvafa\hm\ and more recently \zt)   quantum
mechanics
on the moduli space of instantons plays a crucial r\^ole.

The
four-dimensional version of the WZW
theory described in \avatar\
depends on a choice of
a $2$-form $\omega$. The path-integral of
the gauged \wzwf\ is related to the
four-dimensional Verlinde number
\avatar. If we study the theory in
the $\omega \rightarrow \infty$
limit then
the Verlinde number reduces, formally,
to the symplectic volume of the moduli
space of instantons on a four-manifold.
In some cases this moduli space is a
hyperkahler quotient by the ADHM construction.

If a
type IIA 5brane wraps a K3 surface then the resulting
string may be identified with the heterotic string
\hs\sen.
The natural question then arises as to whether one can
wrap IIA 5branes on ALE spaces to produce other types of
strings.\foot{Such objects have recently appeared in
\douglas.}
 A first objection to this idea is that the infinite volume
of  the ALE space leads to an infinite string tension. However,
it is possible that with a regularized version of the volume,
e.g., that described in this paper other well-defined string
theories can be obtained from wrapped IIA 5branes.

The natural hyperkahler quotients to study in the context of string duality
are
Hitchin spaces.
They occur as spaces of collective coordinates, describing the bound
states of $D2$-branes, wrapping a  curve $\Sigma$ in K3.
It is clear
that a thorough study of these spaces is important in
various aspects of supersymmetric
gauge theories \WitDonagi\donagi.

In all cases we
define the  regularized volumes of these spaces and compute them
using localization techniques.
The localization with respect to different symmetries
produces different formulas, thereby establishing   curious identities.
On this route one may get certain sum rules
for solutions of Bethe Ansatz equations \fadba\fadbai.

The last motivation stems from the recent studies of the
Matrix theory of IIA fivebranes. The Higgs branch of such theory
is described by a two dimensional sigma model with target space being
the moduli space of instantons. The volume of target space
is a characteristic of the ground state wave function. The integrals of
some cohomology classes over the moduli space of instantons like those
which we present below can be translated to give
explicit results for correlations functions
of chiral fields in $(2,0)$ six dimensional
theories in the light-cone description \zt.

The paper is organized as follows. In   section $2$ we review
very briefly the localization approach and describe
the quotients we are going to study. In
section $3$ we present
our definitions of regularized integrals.
Sections $4-7$ are devoted
to explicit examples.

\newsec{Quotients and Localization}

\subsec{K\"ahler and hyperk\"ahler  quotients}

A kahler manifold $X$ is by definition a complex variety with
a hermitian metric $g$, such that the corresponding two-form
$\omega ( \c, \c) = g( I \c , \c)$ is closed. The complex structure
$I$, viewed as the section of ${\rm End} (T^{\scriptscriptstyle \IC}X)$
is covarianly closed $\nabla I = 0$, for the
Levi-Civita connection $\nabla$.
A hyperkahler
manifold $X$ has three covariantly constant complex structures $I, J, K$
which obey quaternionic algebra: $I^{2} = J^{2} = K^{2} = IJK = -1$
which are such that in any of the complex structures  the manifold $X$ is
complex. It follows that $\omega^{r} = g(I \c, \c)$ is the K\"ahler
form w.r. t. complex structure $I$ while
$\omega^{c} = g( J \c , \c) + \sqrt{-1} g (K \c, \c)$ is a closed
$(2,0)$-form i.e.,
 a holomorphic symplectic structure. See \hklr\ for a detailed
introduction to the subject. Whenever a compact
group $K$ acts on a symplectic
manifold $(X, \omega)$ preserving its symplectic structure one may attempt
to define a reduced space.
To this end the existence of an
equivariant
moment map $\mu : X \to \liek^{*}$ is helpful.
The latter is defined as
follows: $d\langle \mu , \xi\rangle   =
- \iota_{V_{\xi}} \omega$ where $\xi \in \liek$ and
$V_{\xi} \in {\rm Vect}(X)$ is the corresponding vector field. In other
words, $\langle \mu, \xi \rangle$ is the hamiltonian generating the action
of $\xi$. The
equivariance
condition means that $\mu (g \c x) = Ad_{g}^{*} \mu (x)$ for any $g \in
K$.
The existence
of $\mu$ is guaranteed in the situation where $X$ is simply-connected
while the equivariance
depends
on the  triviality of a certain cocycle of the Lie algebra
of Hamiltonian vector fields on $X$.

The symplectic quotient is the space $X //G = \mu^{-1}(0)/K$.
In the
case where $K$ acts on kahler manifold preserving both its metric
and complex structure the space $X //K$ is kahler as well.
If the group $K$ acts on a
hyperkahler manifold preserving all its
structures then the moment map is extended to the hyperkahler
moment map: $\vec \mu: X \to \liek^{*} \otimes \IR^{3}$, defined
by
\eqn\hprmm{d \langle \vec\mu , \xi \rangle = - \iota_{V_{\xi}} \vec \omega }
where $\vec\omega \in \Omega^{2}(X) \otimes \IR^{3}$ is the triplet
of symplectic forms $g(I \c, \c), g(J \c, \c), g(K \c, \c)$.
It is a theorem of \hklr\ that the hyperkahler quotient
$X ////K = \vec\mu^{-1}(0)/K$ is itself a
 hyperkahler manifold.

In gauge theories with gauge group $K$,
the quotient $X/K$ arises as (a Higgs branch of)
the moduli-space-of-vacua
limit of bosonic gauge theory (whenever spontaneous symmetry breaking
takes place), the kahler quotient $X//K$ occurs
in the theory with $4$ supercharges, the hyperkahler quotients
appear in the theories with $8$ supercharges.

One is tempted to define ``octonionic quotients'' for the purposes
of the theories with sixteen supercharges, but both the notation
($8$ slashes) and the lack of time make us leave it for future
investigations.

The original space $X$ in gauge theories is often linear. So it is
instructive
to play with linear quotients first.  The $0$ in $\mu^{-1}(0)$ can be
replaced
by the central elements $\zeta$, $\vec \zeta$ in $\lieg^{*},
\lieg^{*}\otimes \IR^{3}$. In gauge theories these are called
Fayet-Illiopoulos
terms.
The quotients $\mu^{-1}(\zeta)/K$ and $\vec\mu^{-1}(\vec\zeta)/K$ will be
denoted
as $\CM(\zeta)$ and $\CM(\vec\zeta)$ when this
will not  lead to confusion.

\subsec{Preliminaries: Hyperkahler representations}

Let $K$ be compact group of  $\dim K =k$,
unitarily represented on
a hermitian vector space $V$ of complex
dimension $\ell$.
  $K$ acts via antihermitian matrices  $T_A$.
In a natural way $V\oplus V^*$ is hyperkahler
and the $K$-action
\eqn\kaction{
\delta_A (z^\alpha ; w_\alpha)=
\{(T_A)^\alpha_{~~ \beta} z^\beta ;
- w_\alpha (T_A)^\alpha_{~~ \beta} \}
}
is $\IH$-linear. This motivates the introduction
of quaternionic coordinates:
\eqn\v{
\eqalign{
X^\alpha  = z^\alpha + J w_\alpha
& = \pmatrix{ z^\alpha & \bar w^\alpha\cr
-w_\alpha & \bar z_\alpha\cr} \cr
\delta_A X^\alpha & = (\tau_A)^\alpha_{~~ \beta} X^\beta\cr}
}
where
\eqn\matrx{
(\tau_A)^\alpha_{~~\beta} =
 \pmatrix{(T_A)^\alpha_{~~ \beta}&0 \cr
0& \bigl[ (T_A)^\alpha_{~ \beta} \bigr]^* \cr}
}

On the linear space we have
Kahler form and holomorphic symplectic form:
\eqn\kps{
\eqalign{
\omega^r & = {i\over 2}\sum [dz^\alpha d\zb_{\alpha} +
dw_\alpha d\wb^{\alpha}] \cr
\omega^c & = \sum dz^\alpha\wedge dw_\alpha\cr}
}
We regard exterior derivatives as fermions:
\eqn\drhm{
d X \equiv
\Psi^\alpha = \pmatrix{ \psi^\alpha & \bar \chi^\alpha\cr
- \chi_\alpha & \bar \psi_\alpha \cr}
}

We can also define exterior differentials transforming
as a triplet under the right action of $SU(2)$. Together
the four derivatives are:
\eqn\trip{
\eqalign{
\delta_n X & = \Psi n \cr
\delta_n X^\dagger & = \bar n \Psi^\dagger\cr}
}
here $n$ is any quaternion.
Now let $\rho$ be antihermitian
of norm 1, so
$\rho = u^\dagger i \sigma_3 u$
for
some unitary $u$. Then the holomorphic exterior
differentials in the direction $\rho$ are:
\eqn\extr{
\eqalign{
\p^{(\rho)} &  = \half (d + \delta_\rho)\cr
\pb^{(\rho)} &  = \half (d - \delta_\rho)\cr}
}
Finally,
the (hyper)-kahler potential is
\eqn\hkpot{
\kappa = \half {\Tr} X^\dagger X
}
and satisfies:
\eqn\hkpoti{
\p^{(\rho)}\pb^{(\rho)}  \kappa = \omega^{(\rho)}
= -\half {\Tr} \rho \Psi^\dagger \Psi
}
giving the Kahler form associated to the
complex structure $\rho$.

The moment map for the $K$-action:
\eqn\mm{
d \vec\mu_A + \iota(V_A)\vec\omega =0
}
is given explicitly by:
\eqn\vi{
\eqalign{
X^\dagger_\alpha (\tau_A)^\alpha_{~ \beta} X^\beta
& = 2 \pmatrix{ \mu_A^r & - \overline{ \mu_A^c} \cr
\mu_A^c & - \mu_A^r\cr} \cr
2 \mu_A^r& = \zb_\alpha (T_A)^\alpha_{~\beta} z^\beta
- w_\alpha (T_A)^\alpha_{~~ \beta}\wb^\beta\in \sqrt{-1} \IR \cr
\mu_A^c & = w_\alpha (T_A)^\alpha_{~\beta} z^\beta \cr}
}

We finally construct the hyperkahler quotient.
Let
${\vec\mu}^{-1}(\vec\zeta ) =
{\mu}_{r}^{-1}(\zeta^r) \cap \mu_{c}^{-1}({\zeta}^c)$
where $\vec\zeta_A$ is nonzero only for central generators
$T_{A}$. Then
\eqn\xvii{\CM(\vec \zeta)   = \vec\mu^{-1}(\vec \zeta) /K }
is of real dimension
$4 \ell - 4 \dim K = 4 (\ell -k)= 4 m$.
As complex manifolds we may write \hklr:
 \eqn\compqut{
\CM(\zeta^c ) = \mu_c^{-1}(\zeta^c) /K_{\IC} }
after deleting the unstable points.

\subsec{Review of CohFT approach and localization}

This section is devoted to a review of mathematical and physical
realizations of integration over the quotients in various
( real, complex, quaternionic and octonionic)  cases.
Consider a manifold $X$ with an action of a group $K$. Suppose
that submanifold $N \subset X$ is $K$-invariant and is acted on by $K$
freely.
The problem we
adress is how to present the  integration over the quotient
$\CM = N/K$ in terms of
the one over $X$. Let $x^{\mu}$ denote the coordinates on $X$.
Their exterior derivatives are denoted as $\psi^{\mu}$.
Differential forms $a \in \Omega^{*}(X)$
on $X$ are  regarded as functions of $(x, \psi)$.
Equivariant forms are   equivariant maps:
\eqn\equfrms{\alpha: \liek \to \Omega^{*}(X)}
of the Lie algebra to the differential forms:
\eqn\equcnd{{\alpha}({\rm Ad}_{g}{\phi}) = g^{*} {\alpha}({\phi})}
where in the right hand side $g$ represents the action of an element
$g \in K$ on $X$ and correspondingly on $\Omega^{*}(X)$. The equivariant
derivative
\eqn\equder{D \equiv d + \iota_{V_{\phi}}}
squares to zero on the space of equivariant forms, which we denote as
$\Omega^{*}_{K}(X)$.

The answer to our problem is positive in the important
case where the $K$-invariant submanifold   $N$ is realized
as a set of zeroes of a $K$ -equivariant
section of a $K$ -equivariant bundle $E \to X$:
$N = s^{-1}(0)$ for $s: X \to E$. In this case it is possible to represent
the integration over $\CM$ in terms of integrals over $X$ and some
auxilliary
supermanifolds. See \atbott\Witr\Witdgt\CMR.
Endow $E$ with a $K$-invariant
metric $(,)$ and let $A$ be the connection which is compatible with it.
One needs
a multiplet $(\chi, H)$ of ``fields'' of opposite statistics
taking values in the sections of $E$. The equivariant derivative
$D$ acts as follows:
\eqn\kcom{\eqalign{D \chi = & H - \psi^{\mu} A_{\mu}  \c \chi\cr
D H = & T_{\phi} \c \chi - \psi^{\mu} A_{\mu} \c H - \psi^{\mu}\psi^{\nu}
\CF_{\mu\nu} \c \chi\cr}}
where $T_{\phi}$ represents the action of $\phi \in \liek$ on the fibers
of $E$ and $\CF_{\mu\nu}$ is the
curvature of $A$. One also needs the ``projection multiplet'' $(\bar\phi,
\eta)$
with values in $\liek$ with the action of $D$ as: $D \bar\phi = \eta,
D \eta = [\phi, \bar\phi]$.

In the situation we are in, namely kahler and hyperkahler quotients
the bundle $E$ is the topologically  trivial bundle
$E = X \times \lieg^{*}$ or $E = X \times \lieg^{*} \otimes \IR^{3}$
respectively. The section $s$ is simply $\mu - \zeta$ and $\vec \mu -
\vec\zeta$
respectively.
Let  $g_{\mu\nu}$ be any $K$-invariant metric on $X$. Define a $\liek^{*}$
valued one-form
$\beta$ as follows: $\beta ( \xi) = \half g(V_{\xi}, \c )$ for any $\xi \in
\liek$.
In coordinates $(x, \psi)$ it is simply : $\beta (\xi) = \half
g_{\mu\nu}  \psi^{\mu} V^{\nu}_{\xi}$.

Equivariant cohomology of $X$ maps to the ordinary cohomology of $\CM$.
Indeed, there is an equivariant inclusion map $i: N \to X$
hence the map $i^{*}: H_{K}^{*}(X) \hookrightarrow H^{*}_{K}(N)$ and by the
assumption that the action of $K$ on $N$ is free  we get an isomorphism
$I : H_{K}^{*}(N) \approx H^{*}({\CM})$.
The cohomology classes of $\CM$ among others contain the characteristic
classes of the $K$ bundle $N \to \CM$ which are
in one-to-one correspondence with the invariant
polynomials $P({\phi})$  on $\liek$. These correspond to the equivariant forms
$P({\phi}) \in \Omega^{0}(X)$.
So let us compute the integrals over $\CM$ of the cohomology classes
which are in the image of the map $I \circ i^{*}$. Let $\alpha_{1},
\ldots, \alpha_{l}$ be the representatives of the classes in
$H_{K}^{*}(X)$, so that $\alpha_{p}(\phi)
\in \Omega^{*}(X)$. Let $\Theta_{p} = I \circ i^{*}\alpha_{p}$.
Putting all things together we may represent
the integrals of such classes over $\CM$ as:
\eqn\mnin{\int_{N/K} \Theta_{1}\wedge  \ldots \wedge \Theta_{l} =
\int_{X}
{{\CD \chi \CD H \CD x \CD \psi \CD \phi \CD \bar\phi \CD
\eta}\over{{\Vol}(K)}}
e^{i D\left( \chi \c s + {\beta}({\bar\phi}) \right)}
\alpha_{1} (\phi) \wedge \ldots \wedge \alpha_{l}({\phi}) }

In case where $X$ is a symplectic manifold then the form
$\alpha =
\omega + \langle \mu, \phi \rangle \in \Omega_{K}^{*}(X)$ is equivariantly
closed and gives rise to a symplectic form $\varpi$ on the reduced
space $X//K$. Similarly, in the hyperkahler case the triplet
of equivariantly closed forms
$\vec \alpha = \vec \omega + \langle \vec\mu, \phi \rangle $ produce
a triplet of kahler forms $\vec\varpi$ on the quotient
$X////K$.

So the symplectic volume of the quotient is\foot{For later convenience
we put an $i$ in front of $\phi$ - a kind of Wick rotation}
\eqn\vlm{\int_{X//K}  e^{\varpi} =
\int_{\liek} {{\CD\phi}\over{{\Vol}(K)}}
\int_{\liek \oplus \Pi\liek \oplus \liek \oplus \Pi \liek} {\CD \chi \CD H
\CD \eta \CD \bar\phi}
\int_{X} e^{\omega + i \langle \mu, \phi \rangle} e^{i \langle H, \mu \rangle
+ i \langle \chi , d\mu \rangle   - D ( \beta (\bar\phi)) }}
The specifics of this case is the fact that the auxilliary fields
$\eta,\chi,H,\bar\phi$ come in a quartet and sometimes can be integrated
out
by introducing the $D$-exact term
$$
e^{itD(\langle \chi, \bar\phi \rangle)}
$$
into the exponent. In the limit $t \to \infty$ this term dominates
over $D\left( \langle \chi, \mu \rangle - \beta(\bar\phi) \right)$
and allows to get rid of the auxilliary multiplet with the result:
\eqn\smvlm{
\int_{X//K} e^{\varpi} \Theta_{1} \ldots
\Theta_{l} \sim \int_{\liek} {{\CD \phi}\over{{\Vol}(K)}}
\int_{X} e^{\omega + i  \langle \mu, \phi \rangle}
\alpha_{1}({\phi}) \wedge \ldots \wedge \alpha_{l}({\phi})
}
where now the integral over $\phi$ should be understood as the contour
one and the choice of contour is the subtle memory of the eliminated
quartet.
See   \Witdgt\kirwan\ for a thorough discussion of these
matters.

Another instance where auxilliary fields come in quartets is
in
hyperkahler
reduction. Indeed, the multiplets $(\eta, \bar\phi) \oplus
(\chi, H)$ take values in
$\liek \oplus \liek \otimes \IR^{3}$ which makes possible to introduce
the ``mass term'' of the form:
$$
e^{it_{1} D(\langle \chi^{r}, \bar\phi \rangle) + it_{2} D
(\langle \chi^{c}, {\bar H}^{c} \rangle
+ \langle \bar\chi^{c}, H^{c} \rangle)}, t_{1},t_{2} \to \infty
$$
By taking $t_{1} \to \infty$ and integrating
out $H^{c}$ first we arrive at the analogue of \smvlm:
\eqn\hpvlm{\eqalign{&
\int_{X////K} e^{\varpi^{r}} \Theta_{1} \wedge \ldots
{\Theta_{l}} = \cr
&
\int_{\liek} {{\CD \phi}\over{{\Vol}(K)}} \int
{{\CD^{2}\chi^{c}}\over{t_{2}^{k}}}
\int_{X} e^{\omega + \langle \mu, \phi \rangle -
{{\vert \mu^{c} \vert^{2}}\over{t_{2}}} + \langle \bar\chi^{c}, d\mu^{c}\rangle
+
 \langle \chi^{c}, d\bar\mu^{c} \rangle + \langle \phi,
[ \bar\chi^{c}, \chi^{c}] \rangle }
\alpha_{1}({\phi}) \wedge \ldots \wedge \alpha_{l}({\phi})  \cr}
}
Had we naively integrated out
$\chi^{c}$  by dropping the $\langle \chi^{c}, d\bar\mu^{c} \rangle$
term we would get an extra determinant
${\Det} ({\rm ad} {\phi})$ which vanishes due to the zero modes of
${\rm ad}({\phi})$. In fact the mass term does not allow to eliminate the
whole of $\chi^{c}, \bar\chi^{c}$ etc. but rather only its $\liek /\liet$
part, where $\liet$ is the Lie algebra of the maximal torus,
corresponding
to $\phi$\foot{There is also a subtlety in rotating $\phi$ to the torus
but it is irrelevant for us here.}.
So we conclude that the  resulting
integral without the octet of auxilliary fields may be ill-defined.
In fact, it might have been ill-defined from the very begining.
We haven't discussed so far the issue of compactness of the
quotient space. As soon as it is compact we expect the manipulations
we performed to be reasonable and leading to the correct answer.
But what if it is not? We shall discuss it in the next section but
here let us proceed assuming that the non-compactness
can be cured in the $D$-invariant manner.

In this case the other side of the story is the possibility
to express the integrals \mnin\vlm\ in terms of
the fixed points of $K$ action on $X$. This feature comes about due
to the presence of the term
$D (g_{\mu\nu} \psi^{\mu} V^{\nu}_{\bar\phi}) = g(V_{\phi}, V_{\bar\phi}) +
\ldots$
in the exponential which we can scale up by any amount we wish due to its
$D$-exactness. Therefore the integrand is peaked near the zeroes of $V$
and can be evaluated using semi-classical approximation in the directions
transverse to the fixed loci. We present the resulting formulae below.

\subsec{Normal form theorems}

Here we consider the hyperkahler integral \vlm\ and treat it
a bit differently. Namely we integrate out $\chi^{r}, \vec H, \bar\phi,
\eta$
and get the representation for the measure using $(x, \psi, \phi,
\lambda \equiv \chi^{c})$ only. This
exercise is a good warm-up example
yet
it is an illustration of the principle of ``killing the quartets.''
Consider the setup:
\eqn\vii{
\matrix{
 \mu_c^{-1}(\zeta^c) &
\hookrightarrow & X=V\oplus V^* \cr
\downarrow p &  & \cr
\CM_\zeta & & \cr}
}
A $\CCK$-equivariant tubular
neighborhood of the level set will be modeled on a
neighborhood of $(\mu^c)^{-1}(\zeta^c)\times\cliek$
which in turn is modelled on a neighborhood of:
\eqn\compnb{
\CM_\zeta \times T^* \CCK}
Let $\varpi_r, \varpi_c, \bar\varpi_c$ be the
three kahler forms on $\CM_\zeta$.
Then in a ${\CCK}$- equivariant neighborhood of
$\mu_c^{-1}(\zeta^c)$
we have
\eqn\viii{
\eqalign{
\omega_c & = p^* (\varpi_c) +
\sum_{A=1}^{\dim K}
d\bigl(\langle \Theta^A_c ,  \mu_A^c\rangle\bigr)
\cr}
}
where $\Theta^A_c$ is a $(1,0)$ connection form for the holomorphic
principal
$\CCK$ bundle $\mu_c^{-1}(\zeta^c)\to \CM_\zeta$. We claim:
\eqn\x{
\eqalign{
p^*\bigg[(\varpi_c \wedge\bar \varpi_c)^m\biggr] d \mu^{Haar}(K)& =
\biggl[ \omega_c\wedge \bar\omega_c\biggr]^\ell
\prod_{A=1}^{\dim K} \delta^{(2)} (\mu_A^c-\zeta_A^c)
\delta(\mu_A^r-\zeta_A^r) \cr
\int \prod_{A=1}^k  d \lambda_A d\bar \lambda_A
\exp\biggl[
\bar \lambda^A \lambda^B \biggl( &
\zb_\alpha  \{ T_B,T_A\}^\alpha_{~\beta} z^\beta +
w_\alpha  \{ T_B,T_A\}^\alpha_{~\beta} \wb^\beta
\biggr)\biggr]\cr}
}
Indeed, the complexified Lie algebra may
be decomposed
as
$$\cliek = \liek\otimes_\IR\IC = \liek \oplus i \liek $$
according to the decomposition of $\CCK$ into
$K \c H$ where $K$ is compact  and $H$ is from the coset $K  \backslash
{\CCK}$.
Decompose the measure as
\eqn\decmsr{
\biggl[ \omega_c\wedge \bar \omega_c\biggr]^\ell
=
p^*\bigg[(\varpi_c \wedge\bar \varpi_c)^m\biggr]
\biggl[\prod_A \mid d \mu^A_c\mid^2\prod_A \mid d\xi^A_c\mid^2
\biggr]
}
The action of the complex transformations on $z,w$ is
\eqn\cplxact{
\eqalign{
\delta_{iA} z^\alpha & = i (T_A)^\alpha_{~~\beta} z^\beta\cr
\delta_{iA} \bar z_\alpha & = i \bar z_\beta (T_A)^\beta_\alpha\cr
\delta_{iA} w_\alpha & = -  i w_\beta (T_A)^\beta_\alpha\cr
\delta_{iA} \bar w^\alpha & = -  i  (T_A)^\alpha_{~~\beta}\bar w^\beta\cr}
}
Evaluating the delta functions we get the ghost
determinant:
\eqn\ghstdet{
\det_{AB}
 \Biggl[{\p\over \p \xi^B} \mu_A^R( \exp(i \xi^A T_A) .(z,w)\Biggr]
}
and using \cplxact\ to compute the variation of
$\mu^r_A $ in the nilpotent directions gives the
formulae. $\spadesuit$

As an alternative proof we could restrict
$\omega_r$ to
$\mu_c=0$ and then  do ordinary Kahler reduction.
The formula \x\ is equivalent to the  case
$t_{2}=0$ of the formula \hpvlm\
which is easily seen by shifting the $\psi$-variables\foot{It has been
remarked to us by A.~Losev in 1995}, see the analogous
manipulations in the section devoted to instanton moduli.

\newsec{Definitions of the volume}

In this section we are going to deal with the non-compactness which didn't
allow us to perform a large $t_{2}$ evaluation of the formula \hpvlm.
Indeed, $\CM(\vec \zeta)$ is in general noncompact, and
of infinite volume in a sense of ordinary
Riemannian geometry.

Suppose that the space $\CM$ is acted on by a group $H$ in
a Hamiltonian way such that
for some $\epsilon \in \lieh$ the Hamiltonian $H_{\epsilon} = \langle
\mu_{h} , \epsilon \rangle$ is sufficiently positive
at infinity. Then the definition is

a.) The Hamiltonian regularized volume is:
\eqn\hmrg{{\Vol}_{\epsilon}(\CM) = \int_{\CM} e^{\varpi -
H_{\epsilon}}}

Consider the case where $X$ is a linear space.
This space always has a $U(1)$ symmetry which commutes with the
action of the group $K$, namely $(z, w) \mapsto (e^{i\theta}z ,
e^{i\theta}w)$.
This action does not preserve the hyperkahler structure but preserves
the kahler form $\omega^{r}$. If $\zeta^{c} = 0$ then this action
descends to the
quotient space $\CM(\zeta)$. In fact, the corresponding Hamiltonian
coincides with the Kahler potential $\kappa$ and it descends
to the quotient even if $\zeta^{c} \neq 0$ since it is $K$ -invariant.
In this case the Hamiltonian
regularized volume is called kahler regularized volume.

Another natural Hamiltonian is $H_{\epsilon} = {\Tr} \{ \mu_A^r, \mu_B^c\}^c $
where $\{, \}^c$ is the Poisson
bracket in the holomorphic symplectic structure. This regularization is
called ghost regularization.
It will in general differ from the kahler regularization
if $K$ is not simple.

We have for the hyperkahler metric:
${\vol}_{g} = ({\varpi^{r}})^{2m} = (\varpi^{c} \wedge
\bar\varpi^{c} )^{m}$ so the $\epsilon \to 0$ limit of \hmrg\ would
give the Riemannian volume of $\CM$ if only it existed.
Quite analogously we can define regularized intersection numbers,
taking the integrals \hpvlm\ and inserting $e^{-H_{\epsilon}}$.
Notice that in order to {\it define} the kahler or
ghost regularized volume we don't actually need
 $\zeta^{c}$ to vanish.

\subsec{Doing integrals straightforwardly}

Before addressing the issue of existence
of Hamiltonians $H_{\epsilon}$ which generate some symmetries of the
quotients $\CM(\vec\zeta)$ we can express
the kahler and ghost regulated volumes as integrals
over the corresponding linear space $X$ using the normal form theorems.
Applying \x\ we have (we changed the notation
$(\phi, H^{c}, {\bar H}^{c})
\to \vec H$):
\eqn\kahlvol{
\eqalign{{\Vol}_{\epsilon} ({\CM} )   =
 \int_{\CM} &
({\varpi}^c \wedge{\bar \varpi}^{c})^m e^{-H_{\epsilon}} =  \cr
 {1\over{{\Vol}(K)}}
\int_{\liek \otimes \IR^{3}}
 d^3 \vec H \int
&
\prod_A d \chi^{c}_A d\bar \chi^{c}_A
\int_{\hat V \oplus \hat V^*}
\prod_{\alpha=1}^\ell d^4 X^\alpha d^4 \Psi^\alpha  \exp\bigl[
I_1 + I_{gh} + I_{reg}  \bigr] \cr}
}
where the action  is given by
\eqn\xviii{
\eqalign{
I_1 & =
i \half {\Tr}  H^A X^\dagger \tau_A X +
\half {\Tr}  \rho \Psi^\dagger \Psi
+  i {\Tr}  H^A  \zeta_A  \cr
I_{gh} & =  \bar \chi^A \chi^B \biggl(
\zb_\alpha  \{ T_B,T_A\}^\alpha_{~\beta} z^\beta +
w_\alpha  \{ T_B,T_A\}^\alpha_{~\beta} \wb^\beta
\biggr)\cr}}
and $$
I_{reg}  = \cases{& $-  \epsilon {\Tr} X^{\dagger} X$,
\quad {\rm kahler} \cr
& $- \epsilon {\Tr} X^{\dagger} \tau_A \tau_A X$, \quad {\rm ghost} \cr}
$$
and
$H^A, \rho,\zeta_A$ are anti hermitian matrices
\eqn\xix{
H^A = \pmatrix{ H^A_r & \overline{H^A_c}\cr
- H^A_c& - H^A_r\cr}
\qquad
\rho = \pmatrix{ \rho_r & \bar \rho_c \cr
- \rho_c & - \rho_r\cr}
}
(so $\phi_r, \rho_r$ are imaginary).
Now, the above expressions are useful
because they are Gaussian in $X,\Psi$, hence
we can do the integrals.
Doing the Gaussian integrals we arrive at the
formula:

\bigskip
\noindent
{\bf Theorem}. The regularized symplectic
volumes of Hyperkahler quotients are given by the formula
\eqn\xii{
\eqalign{
(2 \pi)^{2\ell}
\int_{\liek \otimes \IR^{3}}
 \prod_{A=1}^{\dim K} d^3 {\vec H}_A \int d \chi^{c}_A d\bar \chi^{c}_A
&
{\exp\biggl[ i \Tr \vec H^A \vec \zeta_A \biggr] \over
\det \CN } \cr}
}
\eqn\xxxi{\eqalign{
\CN &= \epsilon R + i \phi^A \otimes T_A + \kappa
\bar \eta^A  \eta^B 1\otimes \{ T_A, T_B\}  \cr}
}
where $R$ is the regulator matrix. It is $1$ for
kahler regularization and $\tau_A^2$ for ghost regularization.

\subsec{Doing integrals using localization}

It turns out that one can considerably simplify the formula \xii\
in the case $\zeta^{c} = 0$. What is
needed is the existence of a
linear action of
$H$ such
that $\mu^{c}$ is transformed non-trivially under its action.
For the action of $H$ to descend down to $\CM$ it is sufficient
for the equation $\mu^{c} =0$ to be invariant under the action
of $H$. By going to the maximal torus we may assume that $H$ is the
torus itself. In fact, we only need
a
one-dimensional subalgebra,
generated by $\epsilon$. With respect to this subalgebra the
trivial bundle $\cliek \times X$ as well
as its sub-bundle $\cliet \times X$ split as sums of line bundles.
We require that the latter has no zero weight components. In this case
there is a
\bigskip
\noindent
{\bf Theorem.} Suppose that $\zeta^{c} =0$, and that the induced
action $\Lambda_{\liek} (\epsilon)$ of $V_{h}({\epsilon})$ on $\cliek$
(which makes $\mu^{c}$ $H$-equivariant)
is such that the operator
$A = \Lambda_{\liek}(\epsilon) + {\rm ad}({\phi})$ is non-degenerate for
generic
$\phi$.
Then
\eqn\xxxx{{\rm Vol}_{\epsilon}(\CM(\zeta^{r}) =
\int_{\liek} {{{\CD} \phi^{r}}\over{{\rm Vol}(K)}}
e^{i \phi^{r} \cdot \zeta^{r}}
{{{\rm Det}_{\liek} \left( \Lambda_{\liek}(\epsilon) + i {\rm ad} \phi^{r}
\right)}\over{
{\rm Det}_{V \oplus V^{*}}
\left( \Lambda_{X}(\epsilon) + i \phi^{r}_{A} T_{A} \right)
}}}
Here we denoted by $\Lambda_{X} ( \epsilon)$
the linear operator on $X$ which is the derivative of $V_{h}(\epsilon)$ at
$x = 0$.
The measure in \xxxx\ is invariant under the adjoint
action of $K$. It implies that the integral in \xxxx\ can be reduced
to the integral over the maximal Cartan subalgebra $\liet \subset \liek$:
\eqn\vvv{{\rm Vol}_{\epsilon}(\zeta^{r}) =
\int_{\liet} {\CD} t  Z_{\epsilon}\bigl[
{\CM}({\zeta^{r}) \bigr] (t)})}
The  measure $Z_{\epsilon}$ is the measure in \xxxx\ evaluated at the
element $t \in \liek$ times the Vandermonde determinant and ${\CD} t$ is the
standard
Euclidean measure.
The proof of   \xxxx\ involves some use of equivariant cohomology
and all the essential ingredients are already explained in the previous
section, except for the remark concerning the non-compactness
of the quotient space. We said there that in order for the manipulations
with ``mass terms'' to make sense some sort of compactness is required.
Now the regularization we choose provides such a compactness in the
sense that the forms we integrate exponentially fall at infinity.
The formalism with equivariant cohomology must be slightly modified
  to take into account the action of the regulating group $H$.
The derivative $D$ gets promoted to $D_{\epsilon} = D
+\iota_{V_{h}(\epsilon)}$.
The troublesome  term $D \langle  \chi^{c}, \bar H^{c}\rangle$
becomes $D_{\epsilon}\langle  \chi^{c}, \bar H^{c}\rangle$ which is
non-degenerate on $\chi$'s by assumptions of the theorem.

Notice that \xxxx\ also suggests the interpretation of the equivariant
volume of $\CM(\vec \zeta)$ as of the generating function of the
interesection numbers of Chern classes of certain
bundles over the ``half'' $M(\zeta)$  of $\CM(\vec \zeta)$
defined by the symplectic
quotient of $V$ by the action of $K$ at the same level
$\zeta^{r}$.\foot{Integrals which are similar to those
in this paper  have been studied in \dpark.}

\subsec{Hyperk\"ahler regularization}

The last definition is the most symmetric in a sense that it preserves
the hyperkahler structure and is the direct analogue of the
``equivariant volumes'' of \givental. Unfortunately, this definition
is rarely useful, because it requires a tri-holomorphic action on $\CM$
of the
torus of dimension ${1\over{4}} {\rm dim}\CM$ which is rather non-generic.
Nevertheless, suppose that $\CM(\vec\zeta)$ is acted on by the torus $\bf T$
 of the
stated
dimension and let $\vec \epsilon \in \liet \otimes \IR^{3}$ (all examples
studied in \gr\ obey this condition).

b.) Hyperkahler regularized volume:
\eqn\hprv{{\Vol}_{\vec \epsilon}\left( \CM (\vec\zeta )\right) =
\int_{\CM (\vec\zeta)} {\vol}_{g} e^{\langle \vec\epsilon, \vec\mu_{t}
\rangle}}

It essentially reduces to the Hamiltonian regularization in case where
$\epsilon \in \liet \otimes ( \IR \subset \IR^{3})$.

\subsec{Example: volume of a point}

The volume of a point computed with the help of Hamiltonian
regularization is instructive.
We take
$X = \IH = \IC^{2} = \{ (b_{+}, b_{-}) \}$ and $K = U(1)$. The
standard
$K$ action is:
$$
(b_+, b_-) \mapsto (e^{i \theta} b_+, e^{-i \theta} b_-)
$$
and the hyperkahler moment map is given by:
\eqn\stm{\vec M = (\vert b_{+} \vert^{2} - \vert b_{-} \vert^{2} )
{\vec e}_{3} + b_{+}b_{-} {\vec e}_{-} + {\overline{b_{+}b_{-}}} {\vec
e}_{+}}
relevant integral involves first:
\eqn\frstint{
\eqalign{
\int d \eta d \bar \eta   \int d^2 b_+ d^2 b_-
\exp[i \vec \phi \cdot \vec M - \epsilon (\mid b_+\mid^2 +
&
\mid b_-\mid^2) + 2 \eta
\bar \eta(\mid b_+\mid^2 + \mid b_-\mid^2)]
\cr
 = \int d \eta d \bar \eta
{1 \over
(\epsilon- 2 \eta\bar \eta - i \vert \vec \phi\vert)
(\epsilon- 2 \eta\bar \eta + i \vert \vec \phi\vert) }
& = {4 \epsilon \over (\epsilon^2 + \vec \phi^2)^2} \cr}}
(To do the gaussian integral it is convenient to use
rotational invariance to put $\phi$ into the 3 direction.)
Finally, one must do the integral:
\eqn\phint{
\int d^3 \vec \phi e^{-i \vec \phi\cdot \zeta}
{4 \epsilon \over (\epsilon^2 + \vec \phi^2)^2}
= e^{-\epsilon\mid \vec \zeta\mid }
}

In this case all regularizations agree and produce the same result
since the reduced space is point. There is not much freedom here!

\newsec{Examples: four dimensional hyperkahler manifolds}

\subsec{Asymptotically locally euclidean spaces}

Consider the famous ALE gravitational instantons. In the $A_{n}$
case the metric on the space $X_{n}(\vec\zeta)$ is given by the
following explicit formula where the space is represented as an
$S^{1}$ fibration over $\IR^{3}$ which has singular fibers over
$n+1$ point $\vec r_i \in \IR^{3}$: \eqn\alemet{ds^{2} =   V d\vec
r^{2} + V^{-1} (d\tau + \omega)^{2}} where $\vec r \in \IR^{3}, V =
V(\vec r), \omega \in \Omega^{1}(\IR^{3}), d\omega = \star_{3} dV$,
and the potential $V$ is given by: \eqn\ptnl{V = \sum_{i=1}^{n+1}
{1\over{\vert \vec r - \vec r_i \vert}} } The moduli of the space
are $\vec\zeta_i = \vec r_{i+1} - \vec r_i$, $i=1,\dots, n$. We now
represent the $A_n$ ALE spaces $X_n(\vec \zeta)$ as a hyperkahler
quotient, following \kronheimer. We take \eqn\stp{\eqalign{ X = &
\prod_{i=0}^{n}\IC^{2} = \{ (b_{i,i+1}, b_{i+1, i} ) \vert i =0,
\ldots n+1 \equiv 0 \}, \cr K= \prod_{i=0}^{n} & U(1)_i / U(1)_d
\cong \prod_{i=1}^{n} U(1)_{i} = \{ e^{i\theta_{l}} \vert l=0,
\ldots n \} / U(1)_{d}, \cr & H = U(1) = \{ e^{i\alpha}\}\cr}} and
the action of the groups is as follows \eqn\gac{\eqalign{ b_{i,i+1}
& \to e^{i(\alpha + \theta_{i} - \theta_{i+1})} b_{i, i+1} \cr
b_{i+1,i} & \to e^{i(\alpha + \theta_{i+1} - \theta_{i})} b_{i, i+1}
\cr}}

The K\"ahler regularized volume can be computed from our generalized
integral formulae above. After some computation the integral can be
reduced to the intuitively pleasing form:
\eqn\volale{{\Vol}_{\epsilon} (X_{n} \left(\zeta \right))=\int_{X_n(\vec \zeta)}
e^{-\epsilon \kappa} \ {\vol}_{g}}
which can be written more explicitly as
\eqn\volaleii{ {\Vol}_{\epsilon} (X_{n} \left(\zeta \right))= (2\pi)  \int_{{\IR}^{3}} d^3 \vec r
\sum_{i=1}^{n+1} {1 \over \vert \vec r - \vec r_i \vert}
e^{-\epsilon \sum_{i=1}^{n+1} \vert \vec r-\vec r_i\vert } .}

This expression is divergent as $\epsilon \to 0$, but it has an
expansion
\eqn\epexp{
 {\Vol}_{\epsilon} (X_{n} \left(\zeta \right)) = {A_{0} \over \epsilon^2} +{
A_{1}\over \epsilon} + A_{2} + \CO(\epsilon) }
 where $A_i$ are the analytic 
functions of the ${\vec r}_j$'s, invariant under the permutations, as well as the diagonal action of the group of Euclidean isometries ${\IE}(3) = {\IR}^{3} \sdtimes O(3)$. By scaling symmetry $A_j$ scales like
$A_j \to \lambda^j A_j$ under $\vec r_j \to \lambda \vec r_j$. Thus, $A_{1} = 0$, $A_{2} \propto \sum_{j} {\vec r}_{j}^{2} - {1\over n+1} \left( \sum_{j} {\vec r}_{j} \right)^{2}$. To fix the proportionality constant, we employ the following trick.  

{}By our definition, the regularized ``volume'' is $A_2$. This
function can be evaluated as follows. First note that the leading
divergence in ${\Vol}_{\epsilon} (X_{n} \left(\zeta \right))$ for $\epsilon\to 0$ comes
from the region $\vec r\to \infty$. The integrand simplifies in this
region and we find
$$
{\Vol}_{\epsilon} \left( X_{n} \left(\zeta \right)\right) = {8\pi^2\over n+1} {1\over \epsilon^2} +
{\CO}({\epsilon})
$$
Next, note that for each $j$, $j=1,\dots, n+1$, the integral
satisfies the differential equation:
\eqn\diffeq{\eqalign{& (\nabla_j^2 - \epsilon^2) {\Vol}_{\epsilon} \left( X_{n} \left(\zeta \right)\right)  = \cr
& \qquad\qquad -8
\pi^2 e^{-\epsilon \sum_{i} \mid \vec r_i - \vec r_j\mid} -
4\pi\epsilon \int d^3 \vec r \ {1\over \vert \vec r- \vec r_j\vert}\
\left( \sum_{i\not=j} {1\over \vert \vec r-\vec r_i\vert} \right) \ e^{-\epsilon
\sum_{i=1}^{n+1} \vert \vec r - \vec r_i\vert} \cr}}
The integral on the right hand side is formally of order $\epsilon$ but due to the 
divergence of the integral as $\epsilon\to 0$ at large $\vec r$ this term is in fact
$$
- 16\pi^2 {n\over n+1} + {\CO} (\epsilon)
$$
Using this and plugging \epexp\ into \diffeq\ we find that $A_1(\vec
r_j)=0$ and that
\eqn\atwo{ A_2= -4 \pi^2 {n \over n+1} \biggl( \sum_j \vec r_j^2  +
 \sum_{i\not=j} \kappa_{ij} \vec r_i\cdot \vec r_j
 \biggr)
 }
where $\kappa_{ij}$ are some constants. Now we use the fact that the answer
must be translation invariant to conclude
\eqn\atwo{ A_{2} = -4 \pi^2   \biggl( \sum_{j=1}^{n+1} \vec r_j^2  -
{1\over n+1} (\sum \vec r_j)^2 \biggr)
 }
 This result can be written more compactly as follows. For each
 component (in $\IR^3$) of $\vec r_j$ form the vector in an auxiliary space
  $\rho = (r_1, r_2, \dots , r_{n+1})$ (where we omit the index in $\IR^3$ for clarity).
  Since \atwo\ is translation
  invariant we can assume that $\sum_i r_i=0$, and therefore
  $\rho$ is a vector in the root space of $A_n$. Note that the
  simple roots satisfy $\alpha_i \cdot \rho = r_{i+1}-r_i =
\zeta_i$, $i=1,\dots, n$.  But that means $\rho= \sum_{i=1}^{n}
\zeta_i \lambda_i$ where $\lambda_i$ are the fundamental weights.
But $\lambda_i \cdot \lambda_j = C^{ij}$, the inverse of the Cartan
matrix for $A_n$, and hence we find
\eqn\cartident{
 \sum_{j=1}^{n+1} \vec r_j^2  -
{1\over n+1} (\sum \vec r_j)^2 =   \sum_{i,j=1}^{n} \vec \zeta_i
C^{ij} \vec \zeta_j }
so we have
\eqn\answer{{\Vol}_{\epsilon} (X_n) = {8\pi^2\over n+1} {1\over
\epsilon^2} - 4\pi^2\sum_{i,j=1}^{n} \vec \zeta_i C^{ij} \vec
\zeta_j + \CO(\epsilon^2)}
and hence we obtain the elegant result for the regularized volume of
the $A_n$ ALE space:
\eqn\finalansw{ - 4\pi^2\sum_{i,j=1}^{n} \vec \zeta_i C^{ij} \vec
\zeta_j}

{\bf Remarks}

\item{1.}  It is worth noting that the answer depends
on $\zeta^r$.
 From the QFT perspective,
our integral is a model for a gauge theory with
complexified gauge group $\CCK$.
Thus, the $\zeta^r$ dependence
of the volumes is an {\it anomaly} in the noncompact
part of the complexified gauge group $\CCK$. If we
simply added ghosts we would expect slice-independence:
that is the whole point of Faddeev-Popov gauge fixing.
In fact, the slice parameter $\zeta^r$ for the principal
$\CCK$ bundle does matter. In order to see this dependence
we must regularize with $\epsilon$. In the $\epsilon\to 0$
limit we discover, contrary to naive expectations,
$\zeta^r$ dependence.

\item{2.}
When $\zeta^c_i=0$ and the points $\vec r_j$  lie on a single line
and we can take $\vec r_j=(0,0,z_j)$. In this case the integral
simplifies to
\eqn\epvolii{ \eqalign{ & = (2\pi)^2 \int dz r dr \sum {1 \over
\sqrt{(z-z_i)^2 + r^2} } e^{-\epsilon \sum \sqrt{(z-z_i)^2 + r^2}}
\cr & = {(2\pi)^2 \over  \epsilon} \int dz e^{-\epsilon \sum \mid
z-z_i\mid} \cr  } }
This is an elementary integral which can be done exactly.  Note that
although we could do the integral explicitly, in precisely
 this case the K\"ahler potential can be viewed as a
Hamiltonian, generating a flow on $X_{n}(\zeta^{r})$. So, we can
apply the Duistermaat-Heckmann theorem. The fixed points are the
intersection points of the spheres, where $b_{i,i+1}= b_{i+1,i}=0$.
Another way to evaluate the integral is by acting with the
differential operators ${d^2\over dz_j^2}$ to extract the order
$\epsilon^0$ term, which is once again quadratic in the $z_i$. In
any case, we confirm the result \answer. A nice interpretation of
the regularized volume is that it is the area of the difference
between the graphs of $H(z) = \sum_i | z - z_i|$ and
$H_0(z)=(n+1)\vert z-z_{cm}\vert$ where $z_{cm} = {1\over n+1} \sum_i z_i$.

\item{3.} Let us compare our K\"ahler regularized volume with
that defined by the hyperkahler regularization of the volume. This
gives
 \eqn\hpvl{\int_{X_{n}(\zeta)} e^{i \vec\epsilon
\c (\vec r-\vec r_{cm} )} {\vol}_{g} = {8 \pi^2\over \epsilon^2}
\sum_i e^{i \vec \epsilon \c (\vec r_i-\vec r_{cm} )}  }
where ${\epsilon} = | {\vec \epsilon} |$, and we have subtracted $\vec r_{cm} := {1\over n+1} \sum \vec r_j$
from the hyperk\"ahler moment map to preserve translation
invariance. If we expand in $\vec \epsilon$, the term of order $
\epsilon^0$ depends on the direction $\hat \epsilon = {1\over {\epsilon}} {\vec \epsilon}$. Therefore, we
average over directions to obtain
\eqn\hkvolreg{ {\Vol}_{\epsilon} \left( X_{n} \left(\zeta \right)\right) = {8\pi^2 (n+1)\over
\epsilon^2} - {4\pi^2\over 3}\sum_{i,j=1}^{n} \vec \zeta_i C^{ij}
\vec \zeta_j  + \cdots }
recovering the previous answer, up to a factor of three. 
It seems that this factor of three has something to do with the hyperk\"ahler moment map having three times more components then the ordinary moment map, and so this should be a universal feature of the hyperk\"ahler versus K\"ahler regularisations. We do not, however, have a good understanding of this fact.

\item{4.}  The hyperk\"ahler regularized volume can also be obtained
via a naive cutoff on the radial integral as follows. We begin with
the naive integral for the volume:
 \eqn\intgrl{ \eqalign{ \int_{X_{n} \left(\zeta \right)} \sqrt{g} d^4 x &= 2 \pi
\int_{{\IR}^{3}}  d^{3} r \sum_{j=1}^{n+1} {1\over \vert \vec r - \vec
r_j\vert } \cr}
 }
Next, we cut off the integral at $\vec r\to \infty$ by putting an
upper bound on the radius. More precisely, in order to preserve
translation invariance we integrate over the region
$$
\vert \vec r - {1\over n+1} \sum \vec r_j\vert < R $$
where $R$ is a cutoff. Next we use the elementary integral
\eqn\xv{ \int_{|\vec r -\vec b| < R} d^3\vec r {1\over |\vec r-\vec
a|} = 2\pi R^2 - {2\pi \over 3}  |\vec a - \vec b |^2. }
In this way we obtain:
\eqn\answerii{ V_R = (2\pi)^2(n+1)R^2 - {(2\pi)^2\over 3}
\sum_{j=1}^{n+1} (\vec r_j -  \vec r_{cm})^2 }
and we recover the regularized volume:
\eqn\answeriii{ - {(2\pi)^2\over 3} \sum_{i,j=1}^{n} C^{ij} \vec
\zeta_i \cdot \vec \zeta_j}
This is not surprising since it is related to the hyperk\"ahler
regularization by a Fourier transform.

\item{5.} A striking aspect of the regularized volumes is that they
are {\it negative definite} functions of the moduli. Blowup up a
singularity   {\it reduces} the volume of the total space. This can
be understood to be a very general feature of blowing up manifolds
in toric geometry: The toric polytope of the blown up space is obtained by removing a part from the toric polytope of the original space. Indeed, this picture leads to an alternative way to
compute these volumes (see the end of remark 2 above).

\item{6.} One can view ALE spaces as   degenerate K3 manifolds. More precisely,
near a point where K3 has a singularity of $A,D,E$ type the relevant
part of K3 looks like the corresponding ALE space. The volume of
K3 is finite as it is compact manifold. Its variation near
the point of degeneration may be studied using our simple
integrals without knowledge of the exact metric on K3. Of course, in
this case the answer essentially is given by the intersection form
so there is no real simplification. But in the subsequent
cases the intersection theory itself  of the quotient is unknown or
cumbersome.

\subsec{Taub-Nut spaces}

 ALE spaces are   special examples of {\it gravitational
instantons} - they solve the four dimensional Einstein
equations in Euclidean signature. Close relatives of ALE
spaces are ALF spaces, or Taub-Nut manifolds.
The metric on Taub-Nut space is of the same form \alemet\ with
$V = 1 +  \sum_{i} {1\over \vert \vec r - \vec r_i \vert }$. Therefore
the naive regularization of the volume will lead to
the same result as in the ALE case. Let us check whether it is true
for the Kahler or similar regularization.
Taub-Nut spaces can also be realized as hyperkahler quotients of
a flat hyperkahler space \gr: one starts with
${\IC}^{2n} \oplus \IR^{3} \times S^{1} =
\{ (z^{l}, w_{l} ) \vert  l = 1, \ldots, n \} \oplus \{ (t , \vec x) \}$
which is acted on by ${U(1)}^{n}$ as follows.
\eqn\ac{\theta_{l} :
(z^{l}, w_{l}) \mapsto (e^{i\theta_{l}}z^{l}, e^{-i\theta_{l}}w_{l}) ;
\quad (t, \vec x) \mapsto (t + \sum_{l} \theta_{l} , \vec x)}
with the moment maps:
\eqn\mmaps{\vec \mu_{l} = (\vert z^{l} \vert^{2} - \vert w_{l} \vert^{2} -
x^{r}, z^{l}w_{l} - x^{c}, {\bar z}^{l}{\bar w}_{l} - {\bar
x}^{c})}
where $\vec x = (x^{r}, x^{c}, {\bar x}^{c})$, $x^{r} \in \IR, x^{c} \in \IC$.
The generic Taub-Nut space $Y_{n} (\vec \zeta)$
is obtained by imposing the constraints:
\eqn\constr{
\vec \mu_{l} = \vec \zeta_{l}}
and solving them modulo \ac.
In the case $\zeta^{c}_{l} = 0$
the regularizing $U(1)$ action can be chosen to act as follows:
\eqn\regact{e^{i\alpha}: (z^{l}, w_{l}) \mapsto (e^{i\alpha} z^{l},
e^{i\alpha}
w_{l}), \quad l > 1; \quad (t, \vec x) \mapsto (t, x^{r},  e^{i\alpha}x^{c})}
with the Hamiltonian $H$:
\eqn\regham{H = \vert x^{c} \vert^{2} +
\sum_{l=1}^{n} \vert z^{l} \vert^{2} + \vert w_{l}
\vert^{2}}
We see that $H$ does not constrain the $t, x^{r}$ directions. Actually,
$t$ is gauged away by \ac\ while its
$x^{r}$ is constrained by \constr. Thus we expect to get a
finite answer for the regularized volume.
The result is identical to \epvolii.

\newsec{Hitchin spaces and Bethe Ansatz}

\subsec{The setup}

Consider a $G= U(N)$ gauge theory on a Riemann surface $\Sigma$ of genus
$g$.
The gauge field $A = A_{z} dz + A_{\bar z} d{\bar z}$ can be thought of
(locally) as a $\lieg$-valued one-form.
The topological sectors in the gauge theory (choices
of a bundle $E$)
are classified by
$c_{1} (E)= {1\over{2\pi i}} \int_{\Sigma} {\Tr} F \in \IZ$ - a magnetic flux
 of the $U(1)$ part of the gauge group.
If we project the abelian part out in order
 to study the $SU(N)/{\IZ}_{N}$ theory,
then there is  $w_{2}(E) \in {\IZ}_{N}$ - a discrete magnetic flux,
corresponding to
$\pi_{1}(G/U(1))$.
Denote the space of all gauge fields in a given topological sector
as ${\CA}_{p}$, $p = \langle w_{2}(E), [\Sigma] \rangle$.
The gauge group $\CG$ acts in ${\CA}_{p}$ in
a standard fashion: $A \mapsto A^{g} = g^{-1} A g + g^{-1}dg$.
The cotangent space $T^{*}{\CA}_{p}$ is the set of pairs $(A, \Phi)$
where $\Phi$ - a Higgs field is a $\lieg$-valued one-form,
more precisely, a section
of the bundle ${\rm ad}(E) \otimes \Omega^{1}({\Sigma})$,
$\Phi = \phi_{z} dz + \phi^{\dagger}_{\bar z}d\bar z$.
The Hitchin equations \hi\hid\ can be thought of the hyperkahler moment map
equations for the natural action $\CG$  on  $T^{*}{\CA}_{p}$:
\eqn\hitcheq{\eqalign{&
\mu^{r} = F_{z\zb} + [ \phi_{z},  \phi^{\dagger}_{\zb}] \cr
& \mu^{c} = \pb \phi_{z}  + [A_{\bar z}, \phi_{z}] \cr}}
The quotient $\CM = {\vec \mu}^{-1}(0)/{\CG}$ is called a Hitchin space.
It is a hyperkahler
manifold. By construction, it has a natural $U(1)$-action \hid:
\eqn\uoact{e^{i\theta} : (A, \phi_{z}, \phi_{\bar z}^{\dagger})
\mapsto (A, e^{i\theta} \phi_{z},
e^{-i\theta}\phi^{\dagger}_{\bar z} )}
This action preserves the form $\omega^{r}$, decsending from:
\eqn\kalrfrm{\omega^{r} = \int_{\Sigma} {\Tr}
\delta A \wedge \delta A +  \delta \Phi \wedge \delta \Phi}
and is generated
by the hamiltonian $H$, descending from
\eqn\hmlt{H =
\Vert \Phi \Vert^{2} =
\int_{\Sigma} {\Tr} \phi \phi^{\dagger} \, d^{2}z}
We define the regularized volume of $\CM$ to be:
\eqn\rglrvlm{V({\epsilon}) = \int_{\CM} \exp ( \varpi^{r} - \epsilon H), \quad
\epsilon > 0}
Since $\CM$ is a quotient by the gauge group one can rewrite the integral
\rglrvlm\ as a partition function of a certain two
dimensional gauge theory, generalizing the one studied in  \Witdgt.
In the following we repeat the analysis of the sections $2,3$ in a gauge
theory
langauge.
Recall the field content of the gauge theory,
computing the volume of the moduli
space of flat connections \Witdgt:
\eqn\fldgt{A, \psi_{A}, \phi}
with the nilpotent supercharge $Q_{0}$:
\eqn\such{Q_{0} A = \psi_{A}, \quad Q_{0} \psi_{A} = d_{A} \phi,
\quad Q_{0} \phi = 0}
which squares to a gauge transformation $Q_{0}^{2} = L_{V(\phi)}$
generated
by $\phi$.
The charge $Q_{0}$ is a scalar, so $\psi_{A}$ is a fermionic one-form
with values in the adjoint, $\phi$ is a scalar with values on the adjoint
(not to be confused with  the  Higgs one-form!).
The action of the theory is
$$
S_{0} = \int_{\Sigma} {\Tr} \phi F + {\half} \psi_{A} \psi_{A}
$$
In our problem, the original set of fields is to
be enlarged as we start with $T^{*}{\CA}_{p}$
rather then ${\CA}_{p}$. We add
\eqn\fldgtt{\Phi, \psi_{\Phi}} to take care of that.
Now we also impose three conditions
rather then one $F= 0$ as it used to be in \Witdgt, so we
need more Lagrange multipliers, or actually two more
$Q_{0}$ - multiplets:
\eqn\fldgttt{H^{c}, \chi^{c}; \quad {\bar H}^{c}, {\bar \chi}^{c},}
The action of $Q_{0}$ gets promoted to
\eqn\sucht{\eqalign{& Q_{0} \Phi = \psi_{\Phi}, \quad Q_{0}
\psi_{\Phi} = [ \phi, \Phi], \cr
& Q_{0} \chi^{c} = H^{c}, \quad Q_{0}H^{c} = [ \phi, H^{c}] \cr}}
and $S_{0}$ generalizes to
$$
{\widetilde S}_{0} = S_{0} + Q ({\CR})
$$
\eqn\gf{\quad {\CR} = \int_{\Sigma} {\Tr} \bigl(  {\Phi} \psi_{\Phi}
+  {\bar \chi^{c}} \mu^{c}  + h.c. \bigr)}
The last modification
has to do with the fact that we have an extra symmetry in the problem,
namely the $U(1)$ of  \uoact.
Since  $Q_{0}$ is the equivariant derivative it is easy
to modify it to take care of \uoact:
$$
Q_{\epsilon} = Q_{0} + \epsilon {\CQ},
$$
where ${\CQ}$ acts as follows:
\eqn\dact{\eqalign{& {\CQ} A = \quad {\CQ} \psi_{A} = \quad {\CQ} \phi = 0 \cr
& {\CQ} \Phi = 0, \quad {\CQ} (\psi_{\Phi})_{z} = \phi_{z},
\quad {\CQ} ( \bar \psi_{\Phi}) = - \phi^{\dagger}_{\bar z} \cr
& {\CQ}  \chi^{c}= {\CQ} \bar \chi^{c} = 0, \quad {\CQ} H^{c} = \chi^{c},
\quad {\CQ} \bar H^{c} = - \bar\chi^{c}\cr}}
Finally, the action of our gauge theory assumes the form:
\eqn\actn{S_{\epsilon} = S_{0} + Q_{\epsilon} ({\CR})}
with $\CR$ still given by \gf.

There are three ways of evaluating the partition function
\eqn\prtnfn{Z({\epsilon}) =
{1\over{{\rm Vol}({\CG})}}\int \CD \phi \CD A  \CD \Phi \CD
\psi_{A} \CD \psi_{\Phi} \CD^{2} {\vec  H} \CD^{2}\chi
e^{-S_{\epsilon}}}

The first and the most direct one consist of
integrating out the Lagrange multiplers $\vec H$,
thus enforcing the moment map equations and thereby
reducing \prtnfn\ to \rglrvlm.

The second approach, similar in spirit to
\Witdgt\BlThlgt\btverlinde\gerasimov,
uses the localization
with respect to the action of $\CG$,
thereby reducing \prtnfn\ to the sum over the fixed points
of the action of $\CT$ - the $T$-valued gauge transformations. This reduces
the theory to the abelian one and will lead,
as we will  see presently,  to the Bethe Ansatz equations.

The third approach uses the localization of
\rglrvlm\ onto the fixed points of the \uoact,
studied (for $G=SU(2)$) by Hitchin in \hid. The comparison of the second
and the third approaches leads to an
 interesting set of identities,  which we
discuss only in one simple situation.

\subsec{Localization  by  the gauge  group.}

Using the $Q_{\epsilon}$
invariance of our theory we modify the action in such a way
that the integral becomes more tractable.
We add our favorite mass term and throw away the terms $Q \chi^{c}\bar
\mu^{c}$. In this way we get a new action
$$
{\widetilde S}_{\epsilon} = S_{0} + Q_{\epsilon} (\int_{\Sigma} {\Tr}
\bigl(  {\Phi} \psi_{\Phi}
+  {\bar \chi^{c}} \lambda^{c}  + h.c. \bigr))
$$
which is quadratic in
$\chi^{c}$-$\lambda^{c}$, $\Phi$-$\psi_{\Phi}$ and essentially
quadratic in $A$-$\psi_{A}$. The evaluation of the partition function
$$
\int \CD(\ldots) e^{- {\widetilde S}_{\epsilon}}
$$
proceeds as follows. Fix a gauge \eqn\gge{\phi = {\rm diag}(l_{1},
\ldots, l_{N}), \quad \sum l_{k} = 0} which corresponds to a
decomposition of the bundle $E$ into a sum of the line bundles
\eqn\dcmp{ E = \bigoplus_{k} L_{k} } with the Chern numbers
\eqn\chn{c_{1}(L_{k}), \quad n_{k} = \int_{\Sigma} c_{1}(L_{k})}
which should obey the following selection rule: \eqn\slrl{\sum_{k}
n_{k} =  w_{2}(E) {\rm mod} N } Then the gauge field $A$
decomposes as $A = A^{ab} + A^{\perp}$, $A^{ab}$ being the $\liet
= Lie(T), T = U(1)^{N}$ gauge field, and $A^{\perp}$ the component
of $A$ in orthogonal complement to $\liet$. The action is
quadratic in $A^{\perp}$-$\psi^{\perp}_{A}$. Integrating out all
these multiplets together with the Faddeev-Popov determinant for
\gge\ we get the ratio of determinants: \eqn\rtio{
{{{\Det}_{\Omega^{0} \otimes \lieg/\liet} ({\rm ad} \phi)}\over{
{\Det}_{\Omega^{1} \otimes \lieg/\liet} ({\rm ad} \phi)}} {{
{\Det}_{\Omega^{0} \otimes \lieg} (\epsilon + i {\rm ad}
\phi)}\over{ {\Det}_{\Omega^{1} \otimes \lieg} (\epsilon + i {\rm
ad} \phi)}}} Here $\lieg$ denotes the bundle $E \otimes E^{*} - I
\equiv \oplus_{i,j} L_{i} \otimes L_{j}^{-1}- I$, $\lieg/\liet$
denotes $\oplus_{i \neq j} L_{i} \otimes L_{j}^{-1}$. As usual,
the determinants naively cancel, but the infinite-dimensionality
of the situation makes them cancel only up to the determinants
acting on the spaces of zero modes, whose dimension is given by
the Riemann-Roch formula: \eqn\rrg{{\rm dim} \Omega^{0} \otimes L
- {\rm dim} \Omega^{1} \otimes L = c_{1}(L) + 1 - g} Note that the
similar situation was encountered in the derivation of Verlinde
formula from the gauged WZW theory, using localisation techniques
\gerasimov\btverlinde, so we can apply the similar
technique\footnote{$^{2006}$}{We thank A.~Gerasimov for the
clarifying discussion on this subject, important for the
remainder of this section.}. In our case there is an
important complication, though.

The ratio of the determinants is easily calculable for the special background $\phi$, namely for the covariantly constant one.
In general, the regularization preserving the $Q_{\epsilon}$-invariance, would produce also the $A^{ab}$ and $\psi^{ab}$ dependent couplings.
The $Q_{\epsilon}$-symmetry fixes them in terms of two functions,
the induced superpotential $W({\phi})$ and the induced background charge $T({\phi})$. These two functions are calculated below. The general technique of calculating effective actions in terms of observables of topological gauge theory, using index theory, is developed in \nikthesis.

Without the detailed knowledge of $W$ and $T$ we can do
the integral over the rest of the fields, namely,
$A^{ab}$ as well as $l_{k}$'s and $\psi^{ab}_{A}$.
The rest of the action together with the induced terms is given by
\eqn\indac{
 \int_{\Sigma}  \Bigl[ \sum_{k} \left( l_{k} + {{\p} W \over {\p} l_{k}} \right) F_{k} + {\half} \sum_{i,j} \left( {\delta}_{ij} + {{\p}^2 W \over {\p} l_{i} {\p} l_{j}}  \right) \psi_{i} \wedge \psi_{j} \Bigr] +
 \int_{\Sigma} T({\vec l}) {1\over 8{\pi}} {\cal R}^{(2)} }
where $\psi^{ab}_{A} = {\rm diag} (\psi_{1}, \ldots, \psi_{N})$, $F_{k}$ is the
curvature, corresponding to the $k$'th entry of $A^{ab}$, and
${\cal R}^{(2)}$ is the two dimensional scalar curvature.
In writing \indac\ we have dropped possible $Q_{\epsilon}$-exact terms, as they do not affect the result. We have \chn:
$$
\int_{\Sigma} F_{k} = 2\pi i n_{k}
$$
Although the connection $A_{k}$ in the bundle with nonvanishing $n_{k}$
is not a one-form on $\Sigma$
the difference of two such connections $A_{k}^{\prime} - A_{k}$ is a one-form
$\alpha_{k}$. Fix a metric on $\Sigma$ and let $h_{k}$ be a harmonic
representative of the cohomology class of $F_{k}$.
Then, for any connection in this
topological sector we may write
$$
F_{k} = h_{k} + d\alpha_{k}
$$
for some one-form $\alpha_{k}$,
defined up to a gauge transformation $\alpha_{k} \mapsto
\alpha_{k} + d\beta_{k}$.
Integrating $\alpha_{k}$ out together with the gauge fixing
forces $l_{k}$ to obey $dl_{k} = 0$. At the same time we can split $\psi_{k} = \psi_{k}^{h} + {\delta}
\psi_{k}$, where $\psi_{k}^{h}$ is a harmonic one-form, and ${\delta}{\psi}_{k}$ is orthogonal to the space $H^{1}$ of harmonic one-forms.
The integral over ${\delta}{\psi}_{k}$ cancels the determinant induced by the integral over ${\alpha}_{k}$.  Therefore, our
integral reduces to the sum over topological sectors
$n_{k}$, the integral over $\psi^{h}_{k}$ and the integral over the constant
diagonal matrices $\phi$ with the measure
\eqn\fnl{\eqalign{& Z_{p}({\epsilon}) =
\sum_{\vec n}
\int_{\vec l}
\exp \bigl(2\pi i {\vec l} \c {\vec n}\bigr) \nu_{\vec l}
({\epsilon}) \ {\kappa}^{g}_{\vec l}({\epsilon})   \cr
&  \nu_{\vec l} ({\epsilon}, {\vec n}, g)  = {\exp} \ \bigl( 2{\pi} i \sum_{k} n_{k} {{\p} W \over {\p} l_{k}} + ( g- 1) T ( {\vec l}) \bigr) = \cr
& \qquad
\ \prod_{i \neq j} (l_{i} - l_{j})^{n_{i} - n_{j} + 1 - g} \prod_{i,j}
\bigl( \epsilon + i (l_{i} - l_{j}) \bigr)^{n_{i} - n_{j} + 1 - g} \cr
& \qquad\quad \vec l = (l_{1}, \ldots, l_{N}), \quad \vec n = (n_{1}, \ldots, n_{N})}}
where
\eqn\kap{{\kappa}_{\vec l}({\epsilon})  = {\rm det}\left( {\delta}_{ij} + {{\p}^{2} W \over {\p}l_{i} {\p} l_{j}} \right)}
is induced by the integral over the fermionic zero modes, $\psi_{k}^{h}$.
Both  the sum and integral are restricted: $\sum l = 0$, $\sum n \equiv p
{\rm mod} N$. We can relax the condition on $n$'s by introducing a
factor
$e^{-2\pi i p \sum l}$ in the integral
and dropping the requirement $\sum l  = 0$.

Finally, we have two options.
Either we  first take the sum over $\vec n$
or we first
 take the integral over $\vec l$.
The summation over $\vec n$ is easily performed if we notice that
\fnl\ can be rewritten as \eqn\rewr{\eqalign{&Z_{p} ({\epsilon}) =
\int_{\vec l} \CD \vec l \quad  e^{-2\pi i p \sum_{k} l_{k}}\,
\nu_{\vec l}^{1-g} \ {\kappa}_{\vec l}({\epsilon})^{g} \
\sum_{\vec n} \prod_{k=1}^{N} e^{2\pi i n_{k} \chi_{k}} \cr &
\chi_{k} =  l_{k} + {{N+1}\over{2}} + {{\p} W \over {\p} l_{k}} =
l_{k} + {{N+1}\over{2}} + {1\over{2\pi i}} \sum_{j \neq k}  {\rm
log} {{\epsilon + i (l_{k} - l_{j})}\over{\epsilon - i (l_{k} -
l_{j})}}  \cr & \quad \nu_{\vec l} = {\epsilon}^{N} \prod_{i < j}
l_{ij}^{2} (\epsilon^{2} + l_{ij}^{2}), \quad l_{ij} = l_{i}-
l_{j}\cr}} The sum over $n_{k}$ leads to the delta function on
$\chi_{k}$ supported at integers, or, in other words, the integral
in \rewr\ localizes onto the solutions of the
equations\footnote{$^{2006}$}{It was noticed in
\gerasimovshatashvili\ that the original version of this paper
missed the $\kappa_{\vec l}({\epsilon})$ factor in \fnl.}:
\eqn\ba{- e^{2\pi i l_{k}} = \prod_{j \neq k} {{ l_{k} - l_{j}- i
\epsilon}\over{l_{k} - l_{j} + i \epsilon}}} A look at \fadba,
Eqs. (228) and  a formula below Eq. (291) there suggests that this
equation \ba\ is nothing but the
 Bethe Ansatz Equation for the Non-linear Schr\"odinger model (NLS) !
Under this identification $\epsilon$ maps to the coupling ${\bf
g}$ of the NLS Hamiltonian: \eqn\nls{H^{NLS} = \int dx \Bigl(
\vert \p_{x} \psi \vert^{2} + {\bf g} (\psi^{*})^{2}(\psi)^{2}
\Bigr)} In the strong NLS coupling limit ${\bf g} \to \infty$,
$\epsilon \to \infty$ and \ba\ turns into $l_{k} \in \IZ +
{{N}\over{2}}$, with the usual fermionic exclusion principle.

The second option is to integrate $\vec l$ out keeping $\vec n$ fixed and then
to sum over $\vec n$. This approach seems to be equivalent to

\subsec{Localization  via $U(1)$ action.}

The fixed points of the
$U(1)$ action \uoact\ on $\CM$ necessarliy
have ${\Tr} \Phi^{r} = 0$, i.e.
they form a submanifold of the {\it nilpotent cone} $\CN \subset \CM$.
Let $\CS_{\alpha}$ be a connected component of the set of fixed points.
As  explained
in \hid\ the bundle $E$ splits into a sum of line bundles:
$$
E = \oplus \CL_{k}
$$
The contribution of $\CS_{\alpha}$ is the standard integral
$$
Z_{\alpha} =
e^{-\epsilon d_{\alpha}}\int_{\CS_{\alpha}} {{e^{\varpi^{r}}}\over{{\rm
Eu}_{\epsilon}({\CN}_{\alpha})}}
$$
where $d_{\alpha}$
is the value of $\vert\vert \Phi \vert\vert^{2}$ on $\CS_{\alpha}$
(it is easy to express it in terms of $c_{1}({\CL}_{k})$), $\CN_{\alpha}$
is the normal bundle to $\CS_{\alpha}$ and ${\rm Eu}_{\epsilon}$
is its equivariant Euler class.
By summing up $Z_{\alpha}$ we get another expression for \rewr\ thus
establishing a curious identity obeyed by solutions of Bethe Ansatz
equations \ba. We hope to elaborate on this point in
future publications.

\newsec{Volumes of the moduli spaces of
instantons}

\subsec{Gauge theory approach}

Let $X$ be a hyperkahler manifold of dimension $4l$ and let $\CA$
denotes the space of gauge fields in some principal $G$-bundle
$E$ over $X$. Let $\vec \omega$ be a triple of K\"ahler forms
on $X$ and $\vec \pi$ be the corresponding triple
of Poisson structures. Then $\CA$ carries the following (formal)
K\"ahler forms:
\eqn\trkfrms{\vec \Omega = \int_{X}\langle \vec \pi,
{\Tr} \delta A \wedge \delta A \rangle {\rm dvol}_{g}}
where ${\rm dvol}_{g}$ is the metric volume form and $\langle , \rangle$
is the usual contraction. They are invariant under the gauge group action:
\eqn\ggru{A \to A^{g} = g^{-1}dg  + g^{-1} A g,}
and the hyperkahler moment map is
\eqn\ggmm{\vec \mu = \langle \vec \pi , F \rangle }
 The hyperk\"ahler reduction of the space $\CA$ by
the action of the gauge group $K$ produces a manifold $\CM$ which is
infinite-dimensional unless $l=1$.

Suppose $X$ is a compact four-dimensional hyperkahler manifold, i.e. $K3$ or
$T^{4}$. Then the result of the hyperkahler reduction of the space
of the gauge fields produces a finite-dimensional hyperkahler
space, which is nothing but the moduli
space $\CM_{k}$ of instantons on $X$ of a given instanton charge $k$.
This space
posesses a natural compactification and has finite volume. It depends on
the volume of $X$. In fact, one can deduce from
\Witfeb\ that
\eqn\volinst{{\rm Vol}({\CM}_{k}) = {1\over{N!}}{\rm Vol}(X)^{N}}
where ${\rm Vol}(X)$ is computed with the help of the Kahler metric
and $N$ is the quarter of the real dimension of $\CM_{k}$.

As mentioned above, one of the motivations for
this work is the desire to understand the
QFT formula for the number of
``four-dimensional conformal blocks'' \avatar:
\eqn\verlfrm{
\sum_i (-1)^i
\dim H^i(\CM^+(c_1,c_2); \CL_\omega)
= \int_{\CM^+(c_1,c_2)} \ch (\CL_\omega) Td(T^{1,0} \CM^+)
}
This is related to the  gauged \wzwf\ theory.
Let us rescale the form
$\omega = k \omega_0$ and take the limit $k\to \infty$.
In that case \verlfrm\ becomes:
$k^{\dim \CM^+/2} \Vol_{\omega_0}(\CM^+) $.
In the \wzwt\ theory this leads to
expressions for the symplectic volume of the
moduli space of flat connections in terms of
$\zeta$-functions \Witdgt.
In the \wzwf\ theory  from the path integral
expression we obtain:
\eqn\limgwzw{
\eqalign{Z_\omega^{GWZW_4}\to k^{\dim \CM^+/2}
\int & {[d A d\psi  d \chi d H d \varphi ] \over  \vol \CG}  \cr
\exp\Biggl\{
-{i \over  4 \pi}
\int_{X_4}  \Tr\biggl[
\bar H^{0,2} F^{2,0} + \varphi \omega_0\wedge F^{1,1} &
 + H^{2,0} F^{0,2}\cr
+
\omega_0 \psi\wedge \bar \psi +   \bar \chi^{0,2}  &D_A \psi +
 \chi^{2,0} \bar D_A \bar \psi \biggr] \Biggr\}  \cr}
}
This theory has been considered
in \park\parki\  (where it was called
holomorphic Yang-Mills theory).
On a hyperkahler manifold
we can identify the $(2,0),(0,2)$ Donaldson fermions
with $\eta$:
\eqn\dndlfr{
\eqalign{
\chi^{2,0} & = \omega^c \eta \cr
\chi^{0,2} & = \bar \omega^c \bar \eta \cr}
}
Integrating out $\psi,
\bar \psi$ from the HYM lagrangian we produce all
parts of the above action for computing volumes
described above.

If the manifold $X$ is non-compact it is necessary to regularize the
volume to get a well-defined result.
One  natural regularization uses the
 K\"ahler potential \maciocia:
\eqn\reglrize{
\kappa \to \int_{X_4} \kappa(x^\mu) \Tr F^2
}
where $\kappa(x^\mu)$ is the Kahler potential of $X$,
equal to $\sum \mid  \vec r - \vec r_i\mid $ for ALE space.

Now suppose that $X$ is $\IR^{4}$ or an ALE manifold.
The theory with \reglrize\ added to
the action reduces, upon localization to
the moduli space of instantons, to the
``K\"ahler regularization'' used in this paper.

In case where $X$ is ALE manifold there is another
natural term to add to the Lagrangian:
$$
\int \kappa_0(x) \Tr F\wedge F +
\lim_{r\to \infty} \int_{S_r^3} \iota({\p \over  \p r}) [\omega
\Tr (\epsilon F)]
$$
where $\kappa_0(x)$ is the K\"ahler potential of
$X$ and $\epsilon$ is a generic diagonal matrix.
As shown by Nakajima, this generates a torus
action on the moduli space of instantons on
$X_n(\vec \zeta)$. The fixed points are
the fully reducible connections, i.e., the
theory can be abelianized. On the other hand, the moduli spaces
of instantons on $\IR^{4}$ and ALE manifolds can be described
using ADHM construction. The latter realizes the moduli space as
a finite-dimensional hyperk\"ahler quotient. We may therefore apply
our techniques directly in finite-dimensional case. We shall do it
in the next section.

\subsec{Volumes of ADHM moduli spaces}

In the case where the space $X$ is non-compact the moduli space of
instantons on $X$ might have a nice finite-dimensional description.
For example, the instantons on $\IR^{4}$ with the gauge group
$U(k)$ are described by the hyperkahler quotient with respect to
the group $K = U(N)$ of the space $V \oplus V^{*}$ where
$V = \liek \oplus \IC^{N} \otimes \IC^{k}$, with $\IC^{N}$ being the
fundamental
representation of $K$. The FI term $\zeta$ vanishes in this case. We
can consider a slightly deformed space with $\zeta^{r} \neq 0$. Applying
\xxxx\ we arrive at the formula for the regularized
volume:
\eqn\instrfvlm{{\rm Vol}_{N,k}(\zeta^{r}, \epsilon) =
2^N \int
\prod_{l=1}^{N}
{{ e^{i\zeta^{r} \phi_{l}}d\phi_{l}}\over{{\epsilon}\left( \phi_l (
\phi_{l} + 2 i\epsilon)\right)^{k}}} \prod_{i < j} {{\phi_{ij}^{2} (\phi_{ij}^2
+ 4
\epsilon^2)}\over{ (\epsilon^{2} + \phi^{2}_{ij})^2}}}
It is interesting to study this integral in large $N$ limit.
We can rewrite it (after a shift: $\phi_l \to \phi_l +  i \epsilon$)
as a partition function of a gas of $N$ particles
on a line with the pair-wise repulsive potential
\eqn\repel{
V^{rep}(x) = {\rm log}
\left( 1 +  {{\epsilon^{2}}\over{x^{2}}} \right)
\left( 1 -  {{3\epsilon^{2}}\over{x^{2} + 4\epsilon^2}} \right)}
in the external potential, attracting to the origin in the limit $\zeta^{r}
=0$:
\eqn\exter{V^{ext}(x) = k {\rm log} (\epsilon^{2} + x^{2})}
Assuming the existence of the density of the eigenvalues in the large
$N$ limit we get the following equation for the critical point
$$
\sum_{i<j} V^{rep}(\phi_{ij})^{\prime} = - V^{ext}(\phi_{i})^{\prime}
\Rightarrow
$$
\eqn\crit{v.p. \int  {\rho(x) dx} V^{rep}(x- y)^{\prime}
=   {{2k}\over{N\epsilon^{2}}} {{y}\over{y^{2} + \epsilon^{2}}}}
This equation can be solved in a standard way using
the Fourier transform. It seems that the non-trivial
limit exists only if $N$ and $k$ are taken to infinity with $k/N$ fixed.

The moduli space of instantons on $\IR^{4}$ is acted on by the
rotation group $Spin(4) = SU(2)_{L} \times SU(2)_{R}$. The Cartan
subalgebra
of its Lie algebra is two dimensional and is spanned by two elements
$(\epsilon_1,
\epsilon_2)$ which generate rotations in two orthogonal planes in $\IR^4$.
The corresponding equivariant volume is a slight modification of
\instrfvlm:
\eqn\eevl{{\Vol}_{\epsilon_1, \epsilon_2} (\zeta^r) = {{(\epsilon_1 +
\epsilon_2)^{N}}\over{\epsilon_{1}^{N}\epsilon_{2}^{N}}}
\int \prod_{l=1}^{N} {{e^{i\zeta^r \phi_l} d\phi_l}\over{\left( \phi_l ( \phi_l
+
\epsilon_1 + \epsilon_2)\right)^k}} \prod_{i \neq j}
{{{\phi}_{ij}({\phi}_{ij} + \epsilon_1 + \epsilon_2) }\over
{({\phi}_{ij}+ \epsilon_1) ({\phi}_{ij} + \epsilon_2)}}}

\noindent{\bf Remark.} In the case $N=1$ $\zeta \neq 0$ the hyperK\"ahler
quotient coincides with Hilbert scheme $\left({\IC}^{2} \right)^{[N]}$
of points on $\IC^{2}$. The integral \eevl\ can be further evaluated
by residues, which turn out to be enumerated by all Young tableaux $Y$ of
length $N$, and the contribution of a given Young tableau is computed
as follows. The tableau encodes the partition of $N = \nu_1 + \ldots +
\nu_l$
with the condition that $\nu_1 \geq \nu_2 \geq \ldots \geq \nu_l$.
Let us enumerate the boxes in the tableau by the pairs of integers
$(m,n)$, where $1 \leq m \leq l$, $1 \leq n \leq \nu_m$.
Then the integrand in \eevl\ has a pole at
$$
\phi_{m,n} = - \epsilon_1 (m-1) - \epsilon_2 (n-1)
$$
Moreover the residue can be rather easily evaluated using the
results of \nakheis.

Another example where the moduli space of instantons
is given by the hyperkahler quotient is the instantons on
ALE spaces. For concreteness we consider the $U(k)$ instantons
on $A_{n-1}$ ALE space. The topology of the gauge bundle is
specified by the vectors $\vec w$ and $\vec v$ which enter
Kronheimer-Nakajima
construction \KN. We present the result.

\eqn\vlmal{\eqalign{
&
{\rm Vol}_{n; k, \vec w, \vec v} (\vec \zeta, \epsilon) = \cr
e^{-\sum_{i} \zeta_{i} v_{i}} & \int
\prod_{i=1}^{n} \Biggl[ \prod_{\alpha = 1}^{v_{i}}
\left(
{{e^{i\zeta_{i} \phi^{\alpha}_{i}}
d\phi^{\alpha}_{i}}\over{(\epsilon^{2}+(\phi^{\alpha}_{i})^{2})^{w_{i}}}}
{{\prod_{\alpha < \beta} (\phi^{\alpha\beta}_{i})^{2}
\left( (\phi^{\alpha\beta}_{i})^{2} + 4 \epsilon^{2} \right) }\over{
\prod_{\beta=1}^{v_{i+1}}\left( (\phi^{\alpha}_{i} -
\phi^{\beta}_{i+1})^{2} + \epsilon^{2}\right)}} \right) \Biggr]
\cr}}
where $\phi^{\alpha\beta}_{i} = \phi^{\alpha}_{i} - \phi^{\beta}_{i}$.

\newsec{Regularized volumes of quiver varieties}

The formulae \vlmal\instrfvlm\ can be readily generalized to
cover the case of general quiver variety (again with the restriction
$\zeta^{c} = 0$) \KN\nakajima. So assume that a quiver $\Gamma$ is given.
Let $\Omega$ denote the set of its oriented edges. There is an involution
$\iota : \Omega \to \Omega$ which sends an oriented edge to the same edge of
opposite orientation.
Let
$Vert$ be the set of its vertices $v$. There are two  maps
$s,t: \Omega \to Vert$, assigning to an edge its begining (source) and the
end (target). To each vertex a hermitian vector space $L_{v}$  and
to every element $\omega$ of $\Omega$ a vector space
 $H_{\omega} = {\rm Hom} (L_{s(\omega)}, L_{t(\omega)})$
are assigned. Let $l_{v} = {\rm dim}L_{v}$. The space
\eqn\bigsp{\CV_{\Gamma} = \bigoplus_{\omega} H_{\omega} }
is naturally acted on by the group
\eqn\biggru{\CG_{\Gamma} = \times_{v} U(L_{v})}
and this action preserves the hyperkahler
structure on $\CV_{\Gamma}$.  The quiver variety $\CX_{\Gamma}(V,
\vec \zeta)$ is defined for  $V \subset Vert$, $\vec \zeta: V \to \IR^{3}$.
It  is the hyperkahler quotient of $\CV_{\Gamma}$ with
respect to the subgroup $\CG_{V, \Gamma}$ of \biggru:
\eqn\subgru{\CG_{V, \Gamma} = \times_{v \in V} U(L_{v})}

Let
\eqn\anw{l^{\vee}_{v} = \sum_{\omega : s(\omega) = v; t(\omega) \in Vert
\backslash V} l_{t(\omega)}}
Applying \xxxx\ we arrive immediately at:
\eqn\vlmqui{\eqalign{& {\rm Vol}_{\epsilon}(\CX_{\Gamma}(V,
\vec \zeta)) = \cr
& e^{-\sum_{v} \zeta_{v}^{r} l_{v}} \int \prod_{v \in V} \prod_{\alpha=1}^{l_{v}} \left(
 {{e^{i\zeta^{r}_{v} \phi^{\alpha}_{v}}
d\phi^{\alpha}_{v}}\over{(\epsilon^{2}
+ (\phi_{v}^{\alpha})^{2})^{l_{v}^{\vee}}}}
{{\prod_{l_{v} \geq \beta > \alpha}
(\phi_{v}^{\alpha\beta})^{2}\left( \epsilon^{2}+
(\phi_{v}^{\alpha\beta})^{2} \right)}\over{
\prod_{\omega \in \Omega: s(\omega) = v, t(\omega) \in V}
\prod_{\beta =1}^{l_{t(\omega)}}
\left( \epsilon^{2} + (\phi^{\alpha}_{v} - \phi^{\beta}_{t(\omega)})^{2}
\right)^{\half}}} \right) \cr}}

This example covers a lot of interesting hyperkahler manifolds,
complex coadjoint orbits of $GL_{N}({\IC})$ among them, for example
$T^{*}{\IC\IP}^{N}$ with Calabi metric and deformed cotangent bundles
to generalized flag varieties (see for example \nakajima).

\newsec{Acknowledgements}

We would like to thank A.~Losev  for many useful discussions.
We also thank for an interesting
discussion
 N.~Hitchin and V.~Guillemin.  The research of G.~Moore
is supported by DOE grant DE-FG02-92ER40704,
and by a Presidential Young Investigator Award; that of S.~Shatashvili,
by DOE grant DE-FG02-92ER40704, by NSF CAREER award and by
OJI award from DOE and by Alfred P.~Sloan foundation.
The research of N.~Nekrasov was supported by Harvard Society of Fellows,
partially by NSF under  grant
PHY-92-18167, partially by RFFI under grant 96-02-18046 and partially
by grant 96-15-96455 for scientific schools.

\listrefs
\bye